\documentclass[nofootinbib,reprint,amsmath,amssymb,aps]{revtex4-1}

\usepackage{lipsum}
\usepackage{graphicx}
\usepackage{xcolor}
\usepackage{hyperref}
\usepackage{subcaption}
\usepackage[normalem]{ulem}
\usepackage{graphicx}
\usepackage{dcolumn}
\usepackage{bm}
\usepackage{placeins}
\usepackage{amssymb}
\usepackage{appendix}
\usepackage[shortlabels]{enumitem}
\definecolor{internationalorange}{rgb}{1.0, 0.31, 0.0}
\usepackage{tikz}
\usetikzlibrary{"arrows", "automata", "backgrounds", "calendar", "chains", "matrix", "mindmap", "patterns", "petri", "shadows", "shapes.geometric", "shapes.misc", "spy", "trees"}
\usetikzlibrary{arrows,shapes}
\usetikzlibrary{trees}
\usetikzlibrary{matrix,arrows} 
\usetikzlibrary{positioning}
\usetikzlibrary{calc,through}
\usetikzlibrary{decorations.pathreplacing}
\usepackage{pgffor}				
\usetikzlibrary{decorations.pathmorphing}	
\usetikzlibrary{decorations.markings}
\tikzset{
    vector/.style={decorate, decoration={snake}, draw},
	provector/.style={decorate, decoration={snake,amplitude=2.5pt}, draw},
	antivector/.style={decorate, decoration={snake,amplitude=-2.5pt}, draw},
    fermion/.style={draw=black, postaction={decorate},
        decoration={markings,mark=at position .55 with {\arrow[draw=black]{>}}}},
    fermioncyan/.style={draw=black, postaction={decorate},
        decoration={markings,mark=at position .55 with {\arrow[draw=cyan]{<}}}},
    fermiondif/.style={draw=black, postaction={decorate},
        decoration={markings,mark=at position .7 with {\arrow[draw=black]{>}}}},
    fermionend/.style={draw=black, postaction={decorate},
        decoration={markings,mark=at position 1 with {\arrow[draw=black]{>}}}},
    fermionuchannel2/.style={draw=black, postaction={decorate},
        decoration={markings,mark=at position .4 with {\arrow[draw=black]{>}}}},
    scalardif/.style={dashed,draw=black, postaction={decorate},
        decoration={markings,mark=at position .7 with {\arrow[draw=black]{>}}}},
    scalarend/.style={dashed,draw=black, postaction={decorate},
        decoration={markings,mark=at position 1 with {\arrow[draw=black]{>}}}},
    fermionbar/.style={draw=black, postaction={decorate},
        decoration={markings,mark=at position .55 with {\arrow[draw=black]{<}}}},
    fermionnoarrow/.style={draw=black},
    gluon/.style={decorate, draw=black,
        decoration={coil,amplitude=4pt, segment length=5pt}},
    scalar/.style={dashed,draw=black, postaction={decorate},
        decoration={markings,mark=at position .55 with {\arrow[draw=black]{>}}}},
    scalarcyan/.style={dashed,draw=black, postaction={decorate},
        decoration={markings,mark=at position .55 with {\arrow[draw=cyan]{>}}}},
    scalaruchannel1/.style={dashed,draw=black, postaction={decorate},
        decoration={markings,mark=at position .7 with {\arrow[draw=black]{>}}}},
                  scalaruchannel2/.style={dashed,draw=black, postaction={decorate},
        decoration={markings,mark=at position .4 with {\arrow[draw=black]{>}}}},
    scalarbar/.style={dashed,draw=black, postaction={decorate},
        decoration={markings,mark=at position .55 with {\arrow[draw=black]{<}}}},
    scalarnoarrow/.style={dashed,draw=black},
    electron/.style={draw=black, postaction={decorate},
        decoration={markings,mark=at position .55 with {\arrow[draw=black]{>}}}},
	bigvector/.style={decorate, decoration={snake,amplitude=4pt}, draw},
}

\tikzstyle{block} = [draw, rectangle, 
    minimum height=3em, minimum width=6em]

\tikzset{
    cross/.pic = {
    \draw[rotate = 45] (-#1,0) -- (#1,0);
    \draw[rotate = 45] (0,-#1) -- (0, #1);
    }
}

\begin{document}

\hspace{5.2in} \mbox{CALT-TH/2022-037}

\title{Axion Detection with Optomechanical Cavities}

\author{Clara Murgui}
\affiliation{Walter Burke Institute for Theoretical Physics, California Institute of Technology, Pasadena, CA 91125, USA}
\author{Yikun Wang}
\affiliation{Walter Burke Institute for Theoretical Physics, California Institute of Technology, Pasadena, CA 91125, USA}
\author{Kathryn M. Zurek}
\affiliation{Walter Burke Institute for Theoretical Physics, California Institute of Technology, Pasadena, CA 91125, USA}

\date{\today}

\phantom{preprint}
\begin{abstract}
We propose a novel technique to search for axions with an optomechanical cavity filled with a material such as superfluid helium.  Axion absorption converts a pump laser photon to a photon plus a phonon. The axion absorption rate is enhanced by the high occupation number of coherent photons or phonons in the cavity, allowing our proposal to largely overcome the extremely small axion coupling.   
The axion mass probed is set by the relative frequency of the photon produced in the final state and the Stokes mode. Because neither the axion mass nor momentum need to be matched to the physical size of the cavity, we can scale up the cavity size while maintaining access to a wide range of axion masses (up to a meV) complementary to other cavity proposals.

\end{abstract}
\maketitle

{\textit{Introduction}}  --- 
In recent years the QCD axion and axion like particles have been the subject of renewed and intense theoretical and experimental interest. 
In addition to being a potential solution to the QCD strong CP problem, the axion is a well-motivated dark matter candidate.  The QCD axion, in particular, has a highly predictive relation between the mass and coupling to gluons and photons that can be exploited in axion searches.
The axion is, however, difficult to detect, due to its extremely small couplings to gluons and photons.  Many experiments, known collectively as haloscopes, take advantage of the axion as dark matter, where a laboratory experiment searches for an axion wind created by the earth's movement through the galactic dark matter halo. While there have been many ideas, to date only electromagnetic cavities whose resonance frequency is tuned to the axion mass (such as the ADMX experiment \cite{ADMX:2018gho}) have been able to probe the QCD axion as dark matter, and only over a narrow mass range from 2.7 - 4.2 $\mu$eV. There have been a number of other broadband and resonant proposals to look for axion dark matter, from ABRACADABRA and DM-Radio (in the $10^{-12}-10^{-6}$ eV mass range)~\cite{Kahn:2016aff,Ouellet:2018beu,Ouellet:2019tlz,Brouwer:2022bwo}, CASPEr ($<10^{-6}$ eV)~\cite{Graham:2013gfa,Budker:2013hfa} and MadMaX ($40-400~\mu$eV) \cite{Caldwell:2016dcw}, though all of these proposals are still in the development phase. 
The vast majority of the  axion space, from $10^{-12}-10$~meV, is still unreached.

\begin{figure*}[t]
\centering
	\includegraphics[width=1\linewidth]{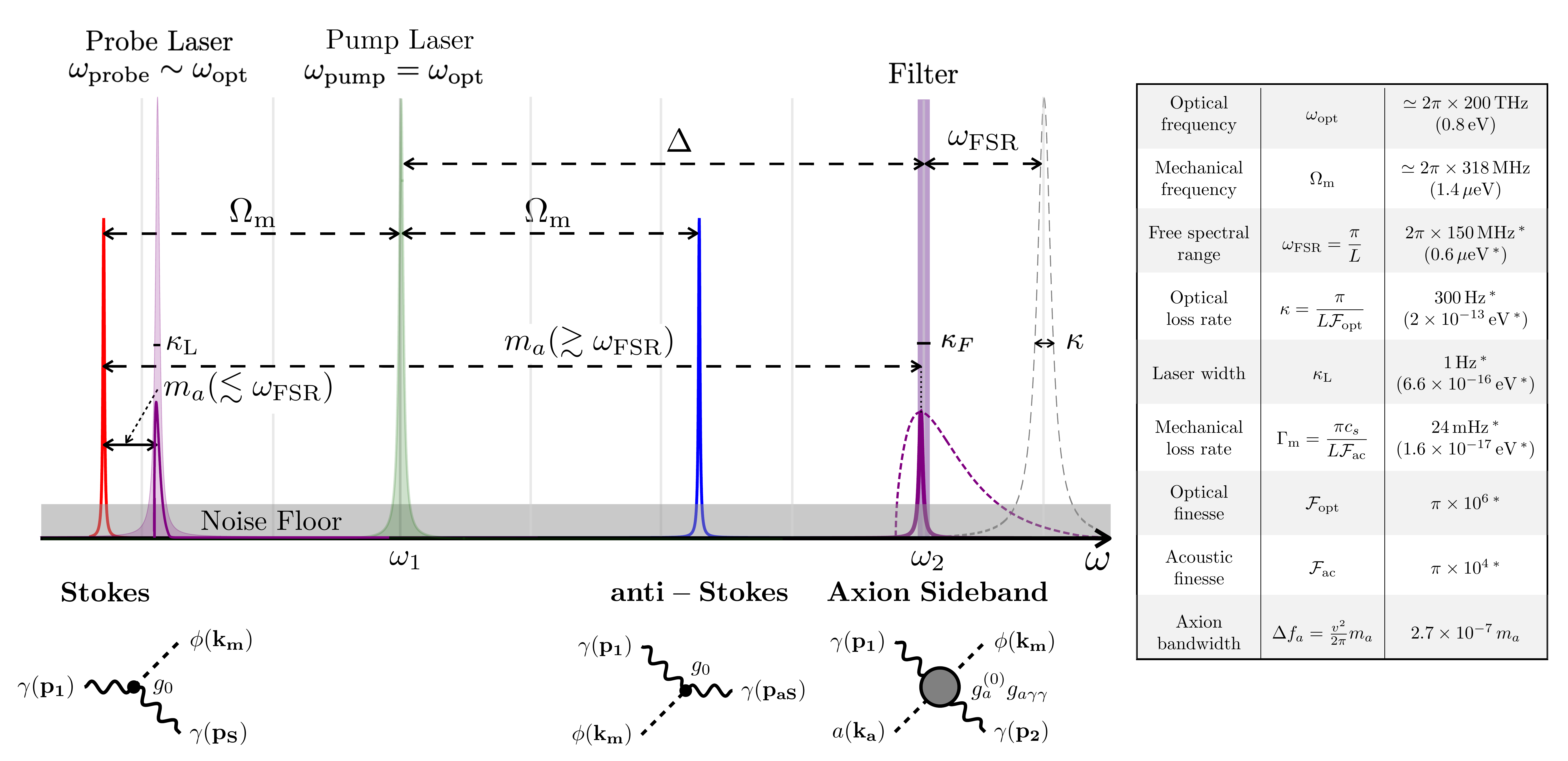}
\caption{Schematic of optical tones and sidebands of the axion signal (purple), and Stokes (red) and anti-Stokes (blue) photons. 
Cavity resonance modes, with a free spectral range $\omega_{\rm FSR}$ and cavity loss rate $\kappa$, are represented by gray vertical lines. The axion mass probed is set by the offset of the photon frequency produced in the final state from the Stokes peak. If interested in axion absorption onto populated photon final states, the axion mass probed is set by the offset of the probe laser (tall purple peak with width $\kappa_L$) from the Stokes peak, convoluted with the energy distribution of the axion (set by the dark matter velocity distribution in the galaxy, shown as a dashed purple peak). For axion absorption onto populated phonon final states, a filter (purple rectangle with width $\kappa_F$) is centered around the cavity density of states, with optical resonance mode $\omega_{n_2}$ instead. There are two different mass regimes shown: (a) the {\it light axion regime} $m_a \lesssim \omega_{\rm FSR}$, where the Stokes and signal are close (in comparison to $\omega_{\rm FSR}$); and (b) the {\it heavy axion regime} $m_a \gtrsim \omega_{\rm FSR}$, where the signal and Stokes photons are separated by more than $\omega_{\rm FSR}$. The table summarizes 
relevant quantities with sample values, evaluated at a benchmark cavity length $L=1\,\text{m}$; those labeled with a star are strongly dependent on the experimental set-up.}
\label{fig:scheme}
\end{figure*}

Against this backdrop, in this {\it letter} we propose a novel mechanism to detect dark matter axions via axion-stimulated phonon emission in a cavity filled with superfluid helium.  In the absence of an axion, a pump laser in the cavity can produce or absorb phonons by a transition to a Stokes or anti-Stokes photon, shown as the large peaks to the left and right of the pump peak operating at $\omega_{\rm opt} $ in Fig.~\eqref{fig:scheme} \cite{aspelmeyer2014cavity,kashkanova2017superfluid,renninger2018bulk,kharel2019high,shkarin2019quantum}.  The pump operates at a cavity mode with harmonic $n$ given by $\omega_{\rm opt} = n \pi / L$.  For example, $
\omega_{\rm opt} = 2\pi \times 200\, {\rm THz}
$ ($\lambda_{\rm opt} = 1.5 \, \mu{\rm m}$) in an existing experiment~\cite{kashkanova2017superfluid,kharel2019high}.  The transition to a Stokes photon happens resonantly when the emission of a phonon at frequency $\Omega_{\rm m}$ redshifts the light, with a pump photon converted, back-to-back, to a Stokes photon.  The phonon absorbs the change in momentum of the photons, $2 \omega_{\rm opt}$, and thus has energy
\begin{equation}
\Omega_{\rm m} \simeq 2\, c_s\,\omega_{\rm opt} \simeq  2\pi \times 318\, {\rm MHz} \simeq 1.36\,\mu\mbox{eV},
\label{eq:mechanicalfrequency}
\end{equation}
where we have approximated the Stokes and pump photons as having the same frequency $\omega_{\rm opt}$.  This is a good approximation since the speed of sound in superfluid helium, $c_s \sim 10^{-6} \ll 1$ such that $\Omega_{\rm m} \ll \omega_{\rm opt}$. (We use natural units $c = \hbar =1$.)

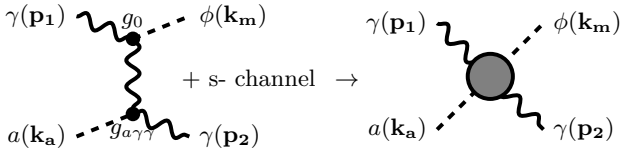
\begin{figure}
    \centering
\begin{equation*}
\begin{gathered}
\begin{tikzpicture}[line width=1.5 pt,node distance=1 cm and 1.5 cm]
\coordinate[label=left:$\gamma (\mathbf{p_1})$](vpump);
\coordinate[right = 0.75 cm of vpump](aux1);
\coordinate[below = 0.3 cm of aux1,label=above:$g_0$](v1);
\coordinate[right = 0.75 cm of aux1,label=right:$\phi (\mathbf{k_m})$](vphonon);
\coordinate[below= 1 cm of v1,label=below:$g_{a\gamma \gamma}$](v2);
\coordinate[below= 0.3 cm of v2](v2aux);
\coordinate[left = 0.75 cm of v2aux,label=left:$a (\mathbf{k_a})$](vaxion);
\coordinate[right = 0.75 cm of v2aux, label=right:$\gamma (\mathbf{p_2})$](vprobe);
\draw[vector](vpump)--(v1);
\draw[vector](v1)--(v2);
\draw[vector](v2)--(vprobe);
\draw[scalarnoarrow](vaxion)--(v2);
\draw[scalarnoarrow](v1)--(vphonon);
\draw[fill=black] (v1) circle (.07cm);
\draw[fill=black] (v2) circle (.07cm);
\end{tikzpicture}
\end{gathered}
\!\!\!\!\!\!\!\!\!\!\!\!\!\!\!\!\!\!\!\!\!\!\!\!
\begin{gathered}
+ \text{ s- channel}
\end{gathered}
\, \, \, \to \!\!\!
\begin{gathered}
\begin{tikzpicture}[line width=1.5 pt,node distance=1 cm and 1.5 cm]
\coordinate[label=left:$\gamma(\mathbf{p_1})$](vpump);
\coordinate[below right = 1 cm of vpump](vcenter);
\coordinate[above right = 1 cm of vcenter,label=right:$\phi(\mathbf{k_m})$](vphonon);
\coordinate[below left = 1 cm of vcenter,label=left:$a(\mathbf{k_a})$](vaxion);
\coordinate[below right = 1 cm of vcenter,label=right:$\gamma (\mathbf{p_2})$](vprobe);
\draw[vector](vpump)--(vcenter);
\draw[scalarnoarrow](vcenter)--(vphonon);
\draw[scalarnoarrow](vaxion)--(vcenter);
\draw[vector](vcenter)--(vprobe);
\draw[fill=gray] (vcenter) circle (.3cm);
\end{tikzpicture}
\end{gathered}
\end{equation*}
    \caption{Diagrams contributing to the effective interaction.}
    \label{fig:diagrams}
\end{figure}

Our idea is to search for the axion by modifying this process through the $2 \rightarrow 2$ axion-stimulated process shown in Fig.~\eqref{fig:diagrams}. 
If the cavity hosts a population of phonons at $\Omega_{\rm m}$, or photons at a frequency $\omega_{\rm opt} - \Omega_{\rm m} + m_a$,  the rate for axion conversion to a phonon plus photon can be quantum mechanically enhanced by the final state coherent population of phonons or photons.  
One of the unique advantages of our approach is that the cavity size need no longer be scaled to match the axion mass; rather the difference of the initial and final state photon momenta can be approximately matched to the phonon momentum, relatively independent of the axion mass. At the same time, the difference in the photon frequencies, the detuning 
between the pump and probe or detection filter,
$\Delta$, is matched to the difference in the phonon frequency and axion mass, $\Delta = m_a - \Omega_{\rm m}$, as shown in Fig.~\eqref{fig:scheme}.
The signal manifests as dark counts from axion-stimulated 
photon
emission when pumping the cavity with 
final-state
phonons, or as shot noise on a probe laser.

The axion induced rate can also be easily estimated.
The power in photons produced from axion absorption, $P_{\rm sig}$, is ostensibly highly suppressed in comparison to an optomechanical cavity with no axion $P_{\rm sig}^{\rm OM}$: 
\begin{eqnarray} \label{eq:}
P_{\rm sig}
 &=&  g_{a\gamma\gamma}^2 \frac{2\rho_a}{m_a^2} \,  P_{\rm sig}^{\rm OM} \\
 &=& 4.7 \times 10^{-44} \left(\frac{g_{a\gamma\gamma}}{10^{-16}\text{ GeV}^{-1}}\right)^2 \left(\frac{\mu{\rm eV}}{m_a} \right)^2 P_{\rm sig}^{\rm OM}, \nonumber
\end{eqnarray}
where $g_{a \gamma \gamma}$ is the axion-photon coupling (defined precisely below), $\rho_a = 0.3 \mbox{ GeV}/\mbox{cm}^3$ the dark matter density and $m_a$ the axion mass. 
We will show that preparing the cavity with a large population of coherent phonons or photons will allow us to overcome the extremely small QCD axion coupling.  We devote the rest of our {\em letter} to computing the axioptomechanical rate, the signal-to-noise ratio, and the theoretical reach of our idea to the dark matter QCD axion or an axion like particle.  In the Supplemental Materials (SM) section, we give more details of our results for the interested reader.

{\textit{Effective Hamiltonian}} --- The Hamiltonian describing the effective interaction shown in Fig.~\eqref{fig:diagrams} is given by
\begin{equation} \label{eq:Heff}
\begin{split}
H_{\rm eff} 
&=-\frac{1}{2}\alpha \, g_{a\gamma\gamma} \, \int d^3 \mathbf{r} \, a(\mathbf{r}) \, n(\mathbf{r}) \, \mathbf{E}(\mathbf{r}) \cdot \mathbf{B}(\mathbf{r}), \\
\end{split}
\end{equation}
where $n(\mathbf{r})$ is the density of the material filling the cavity and $\alpha$ its polarizability.
Quantizing this Hamiltonian, it can be written as a function of the creation and annihilation operators of the mechanical mode ($b^{\dagger},b$, acoustic phonons), optical modes ($a^{\dagger},a$, cavity photons), and the classical axion field amplitude $a(t)$,
\begin{equation} \label{eq:Heffq}
\begin{split}
H_{\rm eff} 
&=  \sum_{i,j,k} \,\,\, \!\!\! g_{a\gamma \gamma} \, g_a^{(0)}  
a(t)
a_{i}  a^{\dagger}_{j} b_{k}^{\dagger}
, 
\end{split}
\end{equation}
where the sum is over all integer mode numbers. $g_a^{(0)}$ is the optomechanical coupling in the presence of the axion source, which depends on the momenta involved in the process (see SM~Sec.~\ref{sec:app1}~for a derivation):
\begin{equation}\label{eq:g0a}
    g_a^{(0)} = \omega_\text{opt} \frac{3}{2} \frac{\varepsilon_r-1}{\varepsilon_r+2}\frac{1}{\varepsilon_r}\sqrt{\frac{|\mathbf{k_{\rm {\bf m}}}|}{2 c_s\rho V_\text{mode} }} a_\text{ovl},
\end{equation}
where 
$V_\text{mode} = L^2 \lambda_\text{opt}/\sqrt{\epsilon_r}$ is the optical mode volume, while $\rho$, $c_s$ and $\varepsilon_r$ are the density, speed of sound and relative permittivity of the material filling the cavity, respectively. 
For superfluid Helium at cryogenic temperatures, $c_s = 238 \text{ m/s}$, $\rho = 145 \text{ kg/m}^3$, and $\varepsilon_r = 1.057$.
The mode overlap factor, 
\begin{equation}
a_{\rm ovl} = \text{sinc}\left( \frac{\pi}{2}\big(n_\text{i} + n_\text{j} - n_{\text{m},k} + \frac{k_a}{\pi/L}\big)\right),
\label{eq:aovl}
\end{equation}
quantifies the overlap between the acoustic and optical modes for a given axion momentum $k_a$ along the cavity axis. Above, $n_i$ and $n_j$ are the natural numbers labelling the resonance optical modes corresponding to the incoming and outgoing photon, respectively, while $n_{{\rm m},k}$ labels the acoustic resonant modes of the outgoing phonon. 
We have assumed that the cavity reduces to a one dimensional configuration along the laser directions, thus the lasers are back-to-back, as we discuss next.

{\textit{Kinematic (phase) matching condition}} ---
The argument of the sinc function in Eq.~\eqref{eq:aovl} is the cavity expression of momentum conservation ({\it i.e.} the cavity mode matching condition), which in the continuum is simply: 
\begin{equation} \label{eq:momentum}
\begin{split}
\quad  {\bf p_1}  +  {\bf k}_{\bf a} &=   {\bf p_2} + {\bf k_m}.
\end{split}
\end{equation}
In general, the axion momentum is very small in comparison to the other momenta, and is irrelevant for the interaction rates.  
For on-shell initial and final states, energy conservation dictates
\begin{equation} \label{eq:}
\begin{split}
    |{\bf p_1}| + m_a & \simeq  |{\bf p_2} | + c_s |{\bf k_m} |, 
\end{split}
\end{equation}
where we approximated the axion energy $\omega_a \simeq m_a$.
Generically, in the regime where $m_a \ll |{\bf p_1}| \sim 0.8 \, {\rm eV}$, and where the phonon energy is small in comparison to the photons (since $c_s \ll 1$), the photons have an equal momentum $|{\bf p_1}| \approx |{\bf p_2}| \approx \omega_{\rm opt}$ and thus are back-to-back. 

In order to meet cavity momentum conservation, and avoid a suppression from $a_{\rm ovl}$ in Eq.~\eqref{eq:aovl}, we require the acoustic and optical modes to be phase-matched. For an axion lighter than the free spectral range (FSR), $m_a \lesssim \pi/L \equiv \omega_\text{FSR}$ ({\em light axion regime}), the axion contribution to the phase matching is negligible, and the condition reduces to the usual requirements for efficient Stokes processes. 
If, however, $m_a \gtrsim \omega_\text{FSR}$ ({\em heavy axion regime}), the phase matching condition is modified to 
\begin{equation}
n_{\rm m} \simeq 2 \, n_\text{1}+ [ m_a /\omega_{\rm FSR}],
\label{eq:sincmatch}
\end{equation}
(see SM Sec.~\ref{sec:appkin} for a detailed derivation). The brackets denote the integer part of their argument.
In this case, the mode overlap factor $a_{\rm ovl}$ of the axion-stimulated versus usual Stokes (without axion) are significantly different.

{\textit{Absorption rate}} ---
According to Fermi's Golden rule, the net rate of absorbing axions, $\Gamma \equiv \Gamma (\gamma + a \to \gamma + \phi) - \Gamma ( \gamma + \phi \to \gamma + a)$, is given by (see~SM~\ref{sec:Srate} for a derivation)
\begin{widetext}
\begin{equation}
\Gamma =(2\pi)  |g_a^{(0)}|^2 \! \left(\frac{2\rho_a}{m_a^2} g_{a\gamma \gamma}^2\right) \! N_{\gamma,\text{pump}}^\text{circ}(\Delta_\text{pump}) \!
\begin{cases}
 N_{\phi}^\text{circ}(\Delta_{\rm m}) \displaystyle \int \!\! d\omega_2 \, B_{m_a}( \omega_2  + \Omega_{\rm m} - \omega_{\rm pump}) L(\omega_2 - \omega_{n_2}, \kappa), \\
N_{\gamma,\text{probe}}^\text{circ}(\Delta_\text{probe}) \displaystyle \int \!\! d\omega_2 \, B_{m_a}( \omega_2  +
\Omega_{n_{\rm m}} - \omega_{\rm pump})  L(\omega_2 - \omega_{\rm probe}, 2\kappa_{\rm L}),
\end{cases}
\label{eq:rate}
\end{equation}
\end{widetext}
in the presence of a pump laser, where the curly bracket indicates that we can either pump the cavity with phonons or a probe laser to quantum mechanically enhance the rate. 
The total rate integrates over the final state photon frequency $\omega_2$, 
where the lineshape is determined by the product of the
density of states of the outgoing photon, parametrized by
the standard Lorentzian distribution, 
\begin{equation}\label{eq:LorentzianStandard}
L( \omega_2 - \omega_{{\rm probe}}, 2 \kappa_\text{L} ) =\frac{1}{\pi} \frac{\kappa_\text{L}}{(\omega_2 - \omega_{\rm probe})^2+\kappa_{\rm L}^2},
\end{equation}
centered at $\omega_{{\rm probe}}$ with a half-width-at-half-maximum $\kappa_{\rm L}$ when the cavity is pumped with a probe laser, or at $\omega_{n_2}$ with full-width-at-half-maximum $\kappa$ when the cavity is pumped with phonons, and the incoming axion flux lineshape, modeled by the Boltzmann distribution
\begin{equation}
\begin{split}
B_{m_a}(\omega_a) = \frac{2}{\sqrt{\pi} \Delta f_a} \sqrt{\frac{\omega_a - m_a}{ \Delta f_a}}
e^{ - \frac{\omega_a - m_a}{ \Delta f_a}}
\Theta(\omega_a - m_a),
\end{split}
\label{eq:Boltzmann}
\end{equation}
centered at $m_a + \Delta f_a/2$ with a bandwidth $\Delta f_a = m_a v^2/2\pi$, with $v = \sqrt{\pi}\times220~$km/s.  We have also used the Fourier transform of $|a(t)|^2$, $|a(\omega_a)|^2= \frac{2 \rho_a}{m_a^2} B(\omega_a)$ in obtaining Eq.~(\ref{eq:rate}). Above, $\omega_{n_2}$ and $\Omega_{n_{\rm m}}$ are the resonance optical and acoustic frequencies, respectively, enhanced by kinematics (see Eq.~\eqref{eq:aovl}). 
The number of circulating photons in the cavity, denoted as $N^\text{circ}_{\gamma,\text{pump}}$ ($N^\text{circ}_{\gamma,\text{probe}}$), is  driven by the pump (probe) laser, given by
\begin{equation} \label{eq:Nphotons}
N_{\gamma,\text{L}}^\text{circ} (\Delta_\text{L}) = N_{\gamma,\text{L}}\,\frac{(\kappa/2)^2 }{\Delta_\text{L}^2 + (\kappa/2)^2},
\end{equation}
where $N_{\gamma,\text{L}}$ is determined by the laser power and frequency, and the cavity loss rate $\kappa = \pi /(\mathcal{F}_{\rm opt}\,L)$:
\begin{equation} \label{eq:Nphotonin}
\begin{split}
&N_{\gamma,\text{L}} 
= \frac{4\,P_{\text{L}}}{\omega_{\rm opt} \kappa} \\
& \simeq 10^{14} \!
\left(\frac{P_\text{L}}{\text{m} \text{W}}\right)  \!
\left(\frac{2\pi\times200\,\text{THz}}{\omega_\text{opt}}\right)
\left( \frac{L}{\rm m} \right)
\left( \frac{\mathcal{F}_{\rm opt}/\pi}{10^6} \right).
\end{split}
\end{equation}
The optical finesse $\mathcal{F}_{\rm opt}$ is determined by the cavity geometry (see~SM~\ref{sec:Srate}).
The Lorentzian in~Eq.~\eqref{eq:Nphotons} with width $\kappa$
accounts for the possible detuning between the laser frequency and the resonance optical mode of the cavity, $\Delta_\text{L}$. Similarly, $N_\phi^\text{circ}$ represents the number of phonons populating the acoustic mode, 
\begin{equation} 
N_{\phi}^\text{circ} (\Delta_{\rm m}) = N_{\phi}\,\frac{(\Gamma_{\rm m}/2)^2 }{\Delta_{\rm m}^2 + (\Gamma_{\rm m}/2)^2},
\label{eq:nphicirc}
\end{equation}
where $\Gamma_{\rm m} $ is the phonon loss rate in the cavity (with $\Gamma_{\rm m} \ll \kappa$), and $\Delta_{\rm m}$ is the detuning between the mechanical drive frequency and the cavity acoustic resonance mode. 
In Eq.~\eqref{eq:rate}, we show the power spectral densities of the outgoing photon in the case $\Delta f_a > \kappa_L$. See SM~\ref{sec:Srate} for the detailed list of cases.

{\textit{Sensitivity}} --- 
The detection modality will depend on whether the cavity is being pumped with phonons or photons for quantum mechanical enhancement in the final state.  If the former, axion absorption can  be detected with a single photon detector; if the latter, the signal is shot noise fluctuations on the probe laser. 
In the case of single photon detection, one must ensure that the pump laser, Stokes and anti-Stokes peaks are filtered in the axion signal frequency window, which means that the tail of the pump must be significantly reduced relative to an input Lorentzian.  
The signal-to-noise ratio (SNR) for a narrowband detection can be estimated as
\begin{equation}
 {\rm SNR} = \Gamma_{\rm sig} \, / \, \Gamma_{\rm back}.
 \end{equation} 
Note that the signal power is penalized if the integration time per scan, $t_\text{int}$, is not longer than the coherence time of the particles participating in the rate, {\it i.e.} axion, cavity photon or laser coherence times. Choosing the integration time to be longer than the laser coherence time, and subsequently longer than the inverse of the optical mode width, $t_\text{int} \gtrsim  \kappa_{\rm L}^{-1} > \kappa^{-1}$, we have
\begin{equation} \label{eq:psig}
\Gamma_\text{sig} \sim \begin{cases}
 \displaystyle t_\text{int} / \tau_a & \text{if } t_{\rm int} \ll \tau_a, \\
 \displaystyle 1 & \text{if } t_{\rm int} \gtrsim \tau_a,
 \end{cases}
 \end{equation}
where the axion coherence time is
\begin{equation}
\tau_a \equiv \Delta f_a^{-1} \!\! = 2.4 \times 10^{-3}\text{s}\left(\frac{\mu \text{eV}}{m_a}\right) \!\! = \frac{(\mu \text{eV}/m_a)}{2.7 \times10^{-13} \text{eV}}.
\end{equation}

The background rate entering in the SNR depends crucially on the detection modality, whether the signal is observed as shot noise fluctuations on top of a probe laser or as counts in a single photon detector. For shot noise fluctuations on a probe laser, the standard deviation of the background reads~\cite{Graham:2013gfa,Foster:2017hbq,Berlin:2019ahk,Dror:2022xpi}
\begin{equation} 
\Gamma_{\rm back} =
\begin{cases}
\displaystyle \kappa \sqrt{N_{\gamma,{\rm bkg}}}/4 & \text{if } t_{\rm int} < \tau_a,\\
 \displaystyle \kappa \sqrt{N_{\gamma,{\rm bkg}}}/(4 \sqrt{N_{\rm meas}}) & \text{if } t_\text{int} > \tau_a,
 \end{cases}
 \label{eq:PsigScaling}
\end{equation}
where $N_{\rm meas} = t_\text{int}\kappa_s$, being $\kappa_s$ the narrowest width in the power spectral density. 
The~SNR agrees with the Dicke radiometer equation~\cite{doi:10.1063/1.1770483} in the long integration time regime. Note that the shot noise fluctuation dominates over other noise sources with an effective noise temperature $T_{\rm n} \ll \sqrt{N_{\gamma,{\rm bkg}}} \,\omega_{\rm opt} \sim 10^4 \sqrt{N_{\gamma,{\rm bkg}}}\, {\rm K}$.
We assume shot noise in the SNR for the curves plotted in Fig.~\eqref{fig:AxionPhoton}(a).

On the other hand, in the case of a single photon counting experiment, the irreducible background is the dark count rate of the detector, $\Gamma_{\rm DCR}$, and the constraint on the axion coupling constant $g_{a \gamma \gamma}$ is computed taking $\Gamma_{\rm sig} t_{\rm int} = 3$, where we take $t_{\rm int} = 1/\Gamma_{\rm DCR}$ in order to maximize the background-free integration time. Note that we are assuming that the tails of the pump laser and mechanical drive, consequently also tails of the Stokes and anti-Stokes sidebands, can be adequately filtered to be a subdominant noise; while this is a significant challenge, it is similar to what is needed for the proposed GQuEST experiment, which aims to observe single signal photons 10 MHz separated from a 10 kW laser having a mHz linewidth~\cite{McCuller:2022hum}. All background photons can be additionally reduced with a polarization filter, noting the different polarization of the photons from the axioptomechanical signal. Even at cryogenic temperatures, the laser induced optical absorption on the cavity mirrors can perturbe and broaden the coherent acoustic modes and drive further thermal phonons. The specific effect depends on the thermal model and material properties, that we delay to a future work. As a crude estimate, the coherent acoustic damping rate should be kept below the optical damping rate. 

We show the theoretical reach, for benchmark parameters with ${\rm SNR}=3$,
to $g_{a\gamma\gamma}$ as a function of the axion mass in Fig.~\eqref{fig:AxionPhoton}.  The coverage in axion mass is achieved via scanning the probe laser in the photon populated final state case, or the frequency of the mechanical drive and the detection filter in the case of a phonon populated final state. In either case, for each measurement, the drives sit at a fixed frequency over the integration time $t_{\rm int}$. 
The reach curve is determined as the $g_{a \gamma \gamma}$ constrained by two subsequent laser (axion energy) widths for light (heavy) axions,
separated by $\delta$, as shown by the inset in the left panel of Fig.~\eqref{fig:AxionPhoton}. The generic scanning strategy consists of matching $\delta$ to the relevant width of $\Gamma_\text{sig}$, up to an order one parameter $\epsilon$, which allows us to enjoy a large portion of its reach.  
Specifics of the scanning strategy and coverage capability can be found in~SM~\ref{sec:e}. 

The reach plots are shown for the two configurations previously introduced: 
\begin{enumerate}[(a)]
\item {\it Populated final state of photons}. In this case, a probe laser is used to scan over different axion masses. The axion signal manifests as shot noise fluctuations on top of the background photons, dominantly from the probe laser, scaling as $(N_{\gamma,\rm probe}^{\rm circ})^{1/2}$. 
As Fig.~\ref{fig:AxionPhoton} shows, for some choices of experimental benchmarks of laser power, cavity length and finesse, such a configuration improves over stellar cooling bounds in the light axion regime; we do not show axion masses below that of a QCD axion with a Planckian decay constant.  
The scanning rate is limited by the laser width $\kappa_L \sim 1\, {\rm Hz} \sim 10^{-15} \text{ eV}$, such that for an integration time $t_{\rm int} = 1\, {\rm s}$ on each mass point one covers approximately 20 neV/year in axion mass, which is sufficient for covering the light axion regime not already constrained by stellar cooling. In the regime where the integration time is longer than the coherent times of the axion and the laser, the dependence on $m_a$ of the axion photon coupling changes from $g_{a \gamma \gamma} \propto m_a^{3/4}$ below 5 neV to $\propto m_a^{3/2}$, with the transition between the two scalings occurring when the axion width is of the same size as the laser width, $\Delta f_a \sim 2 \kappa_L$. 
\item {\it Populated final state of phonons}. The more interesting case is to populate the final state with phonons, where the detection strategy is now to observe single photons from the axion-induced process.  Rather than be limited by laser shot noise, one is limited instead by dark count rates of the single photon detector, as well as photons in the tail of the background tones discussed above.  Here we assume the optimistic scenario in which the background is dominated by the dark count rate of the detector, which is an irreducible source of noise. 
Because the laser tails are more easily filtered with greater separation in frequency from the pump, we focus on the heavy axion case, specifically  $m_a \gtrsim \omega_{\rm FSR}$.   Importantly, in this regime the axion mass becomes significant in the phase matching condition, such that one needs to tune the frequency of the coherent phonons, $\Omega_{\rm m}$, with the axion mass according to Eq.~\eqref{eq:sincmatch}.  
The number of phonons is limited by the density fluctuations in the material, which we require to satisfy $\delta \rho_\text{He} / \rho_\text{He} < 10^{-3}$. 
For more details, we point the reader to SM~\ref{sec:SMphononN}. In this regime, the axion coherence time is always shorter than $t_{\rm int}$, and the reach curve scales as $g_{a\gamma \gamma} \propto m_a^{3/2} (m_a)$ for $\Delta f_a$ broader (narrower) than $\kappa$, as discussed in SM~\ref{sec:e}. For $t_{\rm int}=1\, \text{s}$, one could cover about two orders of magnitude in axion mass over a one year observing time, while for $t_{\rm int} = 10^3 \, \text{s}$, a mass window about 4\% of the mass could be scanned with 10 cavities simultaneously taking data over a year. 
This setup is promising to detect the QCD axion band. Specific choices of inputs and data-taking times can be optimized to scan a specific window, according to the scaling of the parameters and scanning strategies that we have presented here (see SM~\ref{sec:e}). \end{enumerate}

\begin{widetext}
\phantom{A}
\begin{figure}
    \centering
    \includegraphics[width=0.49\linewidth]{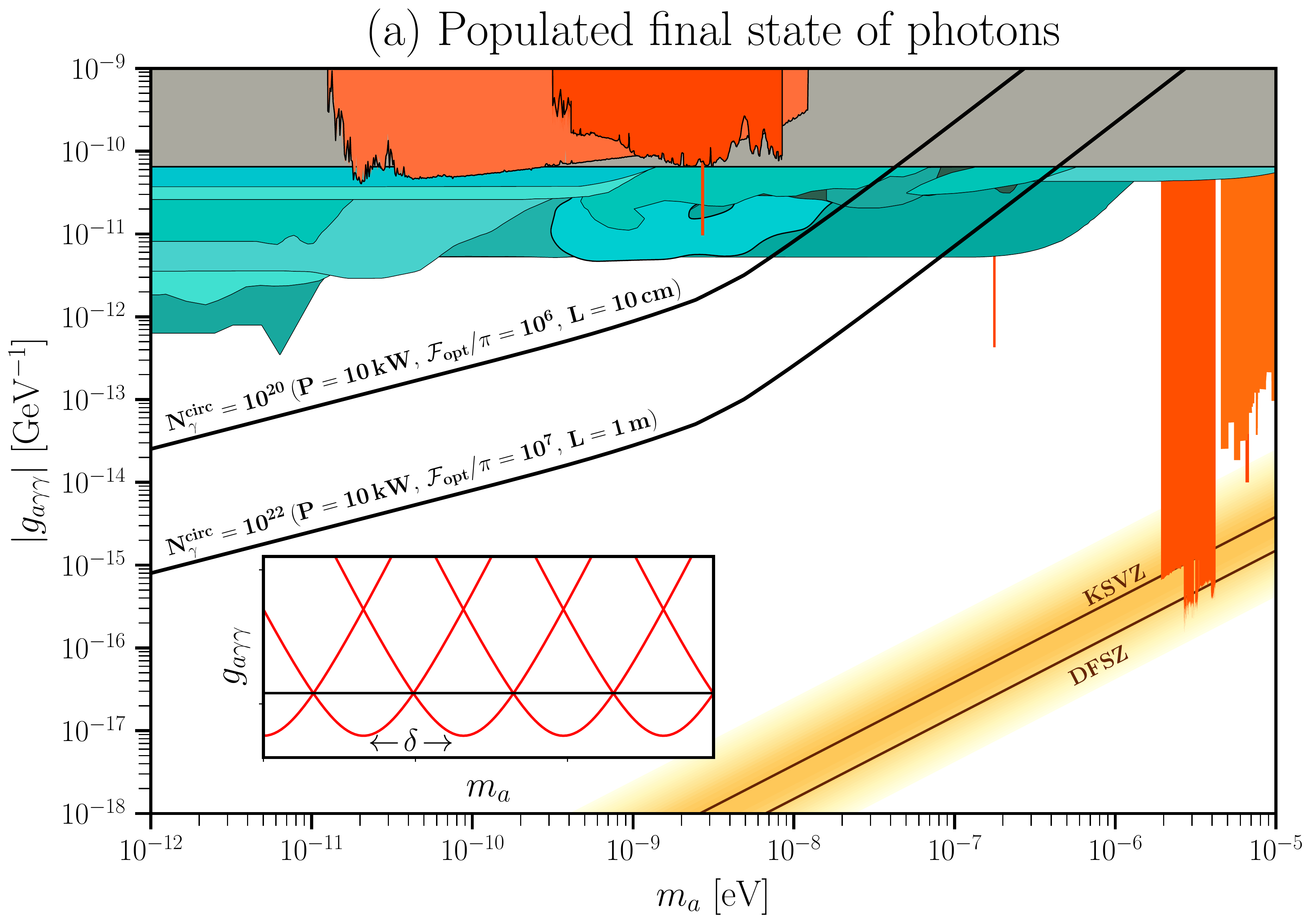}
    \includegraphics[width=0.49\linewidth]{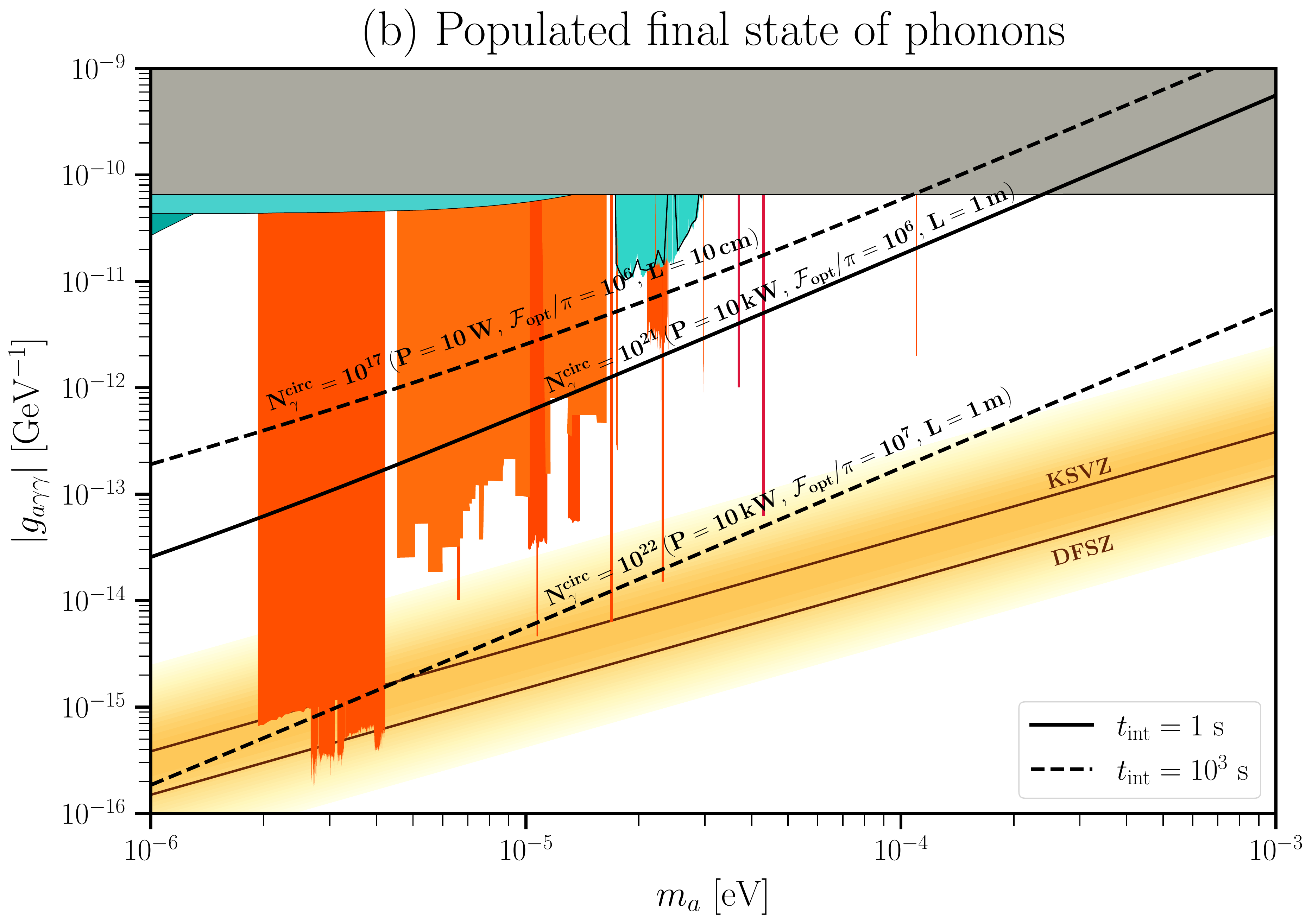} 
    \caption{Theoretical sensitivity reach curves in the $g_{a\gamma\gamma}$-$m_a$ plane for cavity benchmarks with ${\rm SNR}=3$. 
    (a) {\em Populated final state of photons.} The probe laser width is taken to be $\kappa_L = 1$ Hz and the integration time, $t_{\rm int} = 1 \, {\rm s}$. 
The numbers of circulating photons are fixed shown by the labels.    
    (b) {\em Populated final state of phonons.} The phonon number is chosen 
    according to the density perturbation limit of the superfluid Helium.  We show different benchmarks for possible dark count rates, assuming then that $t_{\rm int} \gtrsim \Gamma_{\rm DCR}^{-1}$.
In the sub-panel of plot (a), we show sensitivities (red) for individual measurements with a spacing $\delta$, with the reach curve (black) defined by the intersection between successive measurements.}
    \label{fig:AxionPhoton}
\end{figure}
\end{widetext}

Generically, the sensitivity is dominantly driven by how many photons and phonons can be maintained in the cavity.
High photon numbers will require strong laser power, large optical finesse and long cavities, as shown in Eq.~\eqref{eq:Nphotonin}. 
In the reach plots we fix the number of photons and phonons circulating in the cavity.  The cavity configuration, in general, will need to be tuned for a given axion mass to maintain these populations. For example, the length of the cavity can be tuned to put both the probe and pump lasers on resonance in setup (a), by requiring pump-probe detuning $\Delta = n \, \pi/L = n \, \omega_{\rm FSR}$, where $n$ is an integer. Similarly in setup (b), to scan over a continuous of axion masses by having the pump and mechanical drive on resonance. 
As 
Fig.~\ref{fig:AxionPhoton} highlights, in different axion mass regimes, different cavity \mbox{configurations} may be beneficial for maximizing the number of phonons and photons circulating in the cavity while keeping backgrounds as low as possible.  We postpone a detailed proposal of an experimental strategy for future work~\cite{exp}, restricting ourselves to demonstrating theoretical reach of our idea.  

{\textit{Discussion}} --- 
In this {\it letter}, we have proposed a new idea to search for axions with an optomechanical cavity filled with superfluid helium, hosting coherent optical and acoustic modes driven by a pump laser, and a probe laser or a mechanical drive. We demonstrated theoretical reach to the QCD axion with an $\mathcal{O}(1-10 \text{ m})$ cavity hosting on the order of $10^{20}$ photons and phonons in a coherent state. The axion signal primarily benefits from the large coherent population in the final states that the cavity can host, such that we capitalize on the quantum mechanical enhancement of phonon- or photon-stimulated axion absorption. The latter also provides kinematic matching such that the axion mass decouples from the cavity length, allowing to reach the difficult high axion mass regime with longer cavities that can host more photons and phonons to further amplify the signal.   

While we demonstrated a promising theoretical reach to small axion couplings based on broadly motivated experimental parameters, the optimal experimental route depends on details of the design, including the cavity length, optical finesse, laser power, optical laser width, filtering, the material in the cavity, 
as well as a realistic assessment of reducible backgrounds such as the pump, and (anti-)Stokes tails, thermal and mechanical noises.  We conclude that the promising theoretical reach merits detailed experimental consideration.

\begin{acknowledgments}
We thank Jack Harris and Yogesh Patil for relevant discussions and guidance on the experimental setup of optomechanical cavities. We also thank Dan Carney, Sebastian Ellis, Stefania Gori, Simon Knapen, Nicholas Rodd, and Tanner Trickle for useful feedback. We also thank Asher Berlin and Alex Millar for pointing out the importance of the inverse process $\gamma + \phi \to \gamma + a$, as discussed in the papers \cite{Ioannisian:2017srr,Caputo:2018vmy}. This work is supported by the Quantum Information Science Enabled Discovery (QuantISED) for High Energy Physics (KA2401032), the U.S. Department of Energy, Office of Science, Office of High Energy Physics, under Award Number DE-SC0011632, and by the Walter Burke Institute for Theoretical Physics. The axion limits were plotted using the repository~\cite{AxionLimits}.
\end{acknowledgments}

\bibliography{ref}

\mbox{~}
\clearpage

\widetext
\begin{center}
\textbf{\large Supplemental Materials}
\end{center}
\setcounter{equation}{0}
\setcounter{figure}{0}
\setcounter{table}{0}
\setcounter{section}{0}
\setcounter{page}{1}
\makeatletter
\renewcommand{\theequation}{S\arabic{equation}}
\renewcommand{\thefigure}{S\arabic{figure}}

\appendices

\renewcommand{\thesection}{S.\arabic{section}}

\section{Axioptomechanical coupling}
\label{sec:app1}

The axioptomechanical coupling $g_a^{(0)}$ (see Eq.~\eqref{eq:Heffq}) parametrizes the effective single photon-photon-phonon coupling in the presence of an axion field. We formally derive such an effective coupling and show that 
it can be reduced to the usual optomechanical coupling in the absence of the axion, assuming specific polarizations of the photon fields.

In position space, the number density operator in the effective Hamiltonian described in Eq.~$\eqref{eq:Heff}$ reads $n({\bf r}) = \sum_i \delta^3 \left({\bf r} -{\bf r}_i \right) $. In momentum space, a single phonon excitation state is given by
\begin{equation} \label{eq:}
\begin{split}
| {\bf k} \rangle = \frac{1}{\sqrt{n_0 S({\bf k})}} n_{{\bf k}} | 0 \rangle,
\end{split}
\end{equation}
where $|0\rangle $ is the ground state, $S({\bf k})$ is the static form factor of the target material, and $n_{{\bf k}}$ 
can be written in terms of phonon creation and annihilation operators as
\begin{equation} \label{eq:}
\begin{split}
n_{{\bf k}} = \sqrt{n_0 S ({\bf k})} (b_{-{\bf k}} - b^{\dagger}_{{\bf k}}),
\end{split}
\end{equation}
which is the Fourier transform of the number density operator in position space
\begin{equation} \label{eq:Mq1}
\begin{split}
n({\bf r})  = n_0 + V^{-1/2} \sum_{{\bf k}} e^{i {\bf k} \cdot {\bf r} } n_{{\bf k}},
\end{split}
\end{equation}
where $n_0$ is the average number density of the material.
Optical modes in the cavity correspond to electromagnetic fields with quantized momenta ${\bf p_i}$:
\begin{equation} \label{eq:quantizationp}
\begin{split}
{\bf p_i} = \frac{\pi}{L} \left(i_{x}, i_{y}, i_{z}\right),
\end{split}
\end{equation}
where $i_{x,y,z}$ are integer numbers, and $L$ is the cavity length assumed to be identical in the three dimensions. In our set-up, the lasers drive the optical states to be
one-directional along the cavity axis, defined as~the $z-$direction, {\it i.e.} $i_x, i_y = 0$. Then, optical modes can be labeled by a single integer number $i \equiv i_{z}$, and the momentum $|{\bf p}_i| =  p_{i,z} = |i_z| \, \pi/L$.
Following \cite{scully1999quantum}, the electromagnetic fields can be decomposed as:
\begin{equation} \label{eq:Eq2}
\begin{split}
& {\bf E}({\bf r}, t) 
=
\sum_{i} 
\hat{\epsilon}_{i} \mathcal{E}_{{i}} a_{i}(t) e^{  - i {\bf p}_{i} \cdot {\bf r}} + {\rm h.c.},\\
& {\bf B} ({\bf r},t) = \sum_{i} \frac{ {\bf p}_{i} \times\hat{\epsilon}_{i} }{\omega_{i}} \mathcal{E}_{{i}} a_{i}(t)  e^{ - i {\bf p}_{i} \cdot {\bf r}} + {\rm h.c.} , \\
\end{split}
\end{equation}
where $\hat{\epsilon}_{i}$ is the polarization vector of the mode, and the effective amplitude reads $\mathcal{E}_{{i}} = \sqrt{\frac{ \omega_{i} }{2 \varepsilon_0 V}} $ with $\omega_{i} = |{\bf p}_{i}| $. $a_{i}^{\dagger}$ and $a_{i}$ are the creation and annihilation operators for the corresponding eigenmode, respectively. 

The acoustic modes are also quantized with the same boundary condition. Similar to the optical modes, in one dimension, the acoustic momentum is labeled by a mode number $k$, $|{\bf k}_{{\rm {\bf m}},k}| = k_{{\rm m},k,z} =k \pi/L$, where the subscript `m' stands for {\it mechanical}. Accordingly, we use the mode number $k$ to label the creation and annihilation operator $b_{{\bf k}_{{\rm {\bf m}},k}}^{\dagger} ,b_{\bf -{\bf k}_{{\rm {\bf m}},k}} =b^{\dagger}_k, b_k $ corresponding to such mode.

Following~Eqs.~\eqref{eq:Mq1} 
and \eqref{eq:Eq2}, the interacting Hamiltonian in~Eq.~\eqref{eq:Heff} can be written in momentum space as
\begin{eqnarray} \label{eq:Heffall}
H_{\rm eff} 
 & \supset &
 \frac{\alpha}{2} g_{a\gamma\gamma} a_0 
  \sum_{i,j,k}  
\sqrt{\frac{n_0 S(-{\bf k}_{{\rm {\bf m}},k})}{V}} \mathcal{E}_{i} \mathcal{E}_{j}
 \left( \hat{{\bf p}}_{j} - \hat{{\bf p}}_{i} \right) \cdot (\hat{\mathbf{\epsilon}}_{j} \times \hat{{\bf \epsilon}}^*_{i} )
\int d^3 {\bf r} 
e^{i ( {\bf k}_{{\bf a}} + {\bf p}_{i} - {\bf p}_{j}- {\bf k}_{{\rm {\bf m}},k}  ) \cdot {\bf r} } 
a^{\dagger}_{j}  a^{}_{i} b^\dagger_{k} \nonumber\\
 & =& 
 \frac{\alpha}{2} g_{a\gamma\gamma} a_0 
 \sum_{i,j,k}
\sqrt{\frac{n_0 S(-{\bf k}_{{\rm {\bf m}},k})}{V}} \sqrt{\omega_{i}  \omega_{j}} \frac{1}{2 \varepsilon_0  V} 
\left( \hat{p}_{j} - \hat{p}_{i} \right) \cdot (\hat{\epsilon}_{j} \times \hat{\epsilon}^*_{i} )\,
a^{\dagger}_{j}  a^{}_{i} b^\dagger_{k} \nonumber \\
&&\quad \times
\begin{cases}
\quad (2\pi)^3\delta^3 ({\bf k}_{{\bf a}} + {\bf p}_{i} - {\bf p}_{j}- {\bf k}_{{\rm {\bf m}},k}  ) ,\quad & {\rm full~space}, \\
\displaystyle \prod_{d = x,y,z} \!\!\!L_d \, {\rm sinc} \left(\frac{L_d}{2} \left( k_{a,d} + p_{{i},d} - p_{{j},d} - k_{{\rm m},k,d} \right)  \right), \quad & {\rm cavity~}(L_x,L_y,L_z),
\end{cases}
\end{eqnarray}
where we have rewritten the axion field $a({\bf r}, t) = a_0 (t) e^{i  {\bf k_{\rm a}} \cdot {\bf r} } + $h.c.~with the simplification $\omega_a \sim m_a$. Kinematics in the regime where $m_a \ll \omega_{i,j}$ requires the initial and final photons to be back-to-back (see discussion in the following section~\ref{sec:appkin}), {\it i.e.}~$\hat{{\bf p}}_{j} = - \hat{{\bf p}}_{i} $, which is further confirmed by the above relation as the coefficient is proportional to $\hat{{\bf p}}_{j} - \hat{{\bf p}}_{i} $. From the polarization vector factor, we observe that for circular polarized photons, the initial and final optical modes should have opposite handedness for the polarization cross product not to vanish, such that
\begin{equation}
\begin{split}
\left( \hat{{\bf p}}_{j} - \hat{{\bf p}}_{i} \right) \cdot (\hat{{\bf \epsilon}}_{j} \times \hat{\epsilon}^*_{i} ) = \frac{2}{\varepsilon_{r}},
\end{split}
\end{equation}
where the relative permittivity comes from the normalization of the polarization vector in the medium.
Matching into the effective Hamiltonian in Eq.~\eqref{eq:Heffq}, the momentum dependent axioptomechanical coupling reads
\begin{equation} \label{eq:g0D1}
\begin{split}
 g_a^{(0)}  ({\bf p}_{i},{\bf p}_{j},{\bf k}_{{\rm {\bf m}},k} ,{\bf k_{\rm {\bf a}}}) =    
\sqrt{\frac{ S(-{\bf k}_{{\rm {\bf m}},k})}{n_0 V}} \sqrt{\omega_{i} \omega_{j}} \frac{n_0 \alpha }{ 2\varepsilon_0 \varepsilon_r } 
\begin{cases}
\displaystyle \frac{(2\pi)^3}{ V} \delta^3 ({\bf k}_{{\bf a}} +{\bf p}_{i} - {\bf p}_{j} - {\bf k}_{{\rm {\bf m}},k}  ) ,\quad & {\rm full~space}, \\
\displaystyle \frac{1}{ V} \!\!\! \prod_{d = x,y,z} \!\!\! L_d \, {\rm sinc} \phi_d , \quad  &{\rm cavity~}(L_x,L_y,L_z),
\end{cases}
\\
\end{split}
\end{equation}
where the phases $\phi_d \equiv \frac{L_d}{2} \, \left( k_{a,d} + p_{{i},d} - p_{{j},d} - k_{{\rm m},k,d} \right)$ for $d = x,y,z$ in the cavity setup. Note that momentum conservation in the cavity is represented by a mode overlap factor, $a_\text{ovl}$,
\begin{equation}
\begin{split}
 \frac{(2\pi)^3}{ V} \delta^3 ({\bf k}_{{\bf a}} + {\bf p}_{i} - {\bf p}_{j}- {\bf k}_{{\rm {\bf m}},k}  ) \quad \rightarrow \quad a_{\rm ovl} \equiv
\frac{1}{ V}  \!\!\! \prod_{d = x,y,z} \!\!\! L_d  \,{\rm sinc} \phi_d,
\end{split}
\label{eq:Vdelta}
\end{equation}
which will define the kinematic/phase matching conditions, discussed in the next section. 

In the linear dispersion regime, the static structure function reads $S({\bf k}_{{\rm {\bf m}},k}) = |{\bf k}_{{\rm {\bf m}},k}|/(2 m_{0} c_s)$, where $m_0$ is the nucleon mass of the target. According to the Clausius-Mossotti relation, the polarizability of the medium is related to the relative permittivity as
\begin{equation}
\begin{split}
 \frac{n_0 \alpha}{2 \varepsilon_0 \varepsilon_r } = \frac{3}{2} \frac{\varepsilon_r-1}{\varepsilon_r+2}\frac{1}{\varepsilon_r}.
\end{split}
\label{eq:}
\end{equation}
Incorporating the overlap factor into the axioptomechanical coupling, we finally have
\begin{equation} \label{eq:g0D2}
\begin{split}
g_a^{(0)} 
& = \sqrt{ \omega_{i} \omega_{j} } \sqrt{\frac{|{\bf k}_{{\rm {\bf m}},k}| }{2   c_s\rho V_{\rm mode} }} 
\frac{3}{2} \frac{\varepsilon_r-1}{\varepsilon_r+2}\frac{1}{\varepsilon_r}
a_{\rm ovl}, \\
\end{split}
\end{equation}
where $\rho_0 = m_0 n_0$ is the mass density of the target. Note that the volume factor should be the mode volume as the electric and magnetic fields in~Eq.~\eqref{eq:Eq2} are only non-zero within the volume of the optical mode, $V_\text{mode} = L^2 \lambda_\text{opt} / \sqrt{\epsilon_r}$.

\section{Phase matching condition and the kinematics}
\label{sec:appkin}
In this section, we derive the kinematics and the phase matching condition of the axion absorption process, {\it i.e.}~
\begin{equation} \label{eq:}
\begin{split}
\gamma_{1}\, ({\bf p_{\bf 1}})\, + a \,({\bf k_{\rm {\bf a}}} )\, \rightarrow  \gamma_{2} \,({\bf p_{\bf 2}})\, + \phi\, ({\bf k_{\rm {\bf m}}}) ,
\end{split}
\end{equation}
and specify the axion mass regime for the conditions to be valid.
The momentum and energy conservation read
\begin{equation} \label{eq:cons-all}
\begin{split}
{\bf p_{\bf 1}} + {\bf k_{\rm {\bf a}}} &= {\bf p_{\bf 2}} + {\bf k_{\rm {\bf m}}} \quad \rightarrow \quad
\omega_1 \, \hat{{\bf n}}_{\bf 1} +   m_a v \, \hat{{\bf n}}_{\bf a} = \omega_2\, \hat{{\bf n}}_{\bf 2} +  \frac{\Omega_{\rm m}}{c_s}\, \hat{{\bf n}}_{\rm {\bf m}},\\
\omega_1 + m_a &= \omega_2 + \Omega_{\rm m},
\end{split}
\end{equation}
where we have considered generic directions of the initial and final states, characterized by the unit vectors $\hat{{\bf n}}_{\bf i}$ with $i = 1,2$, a, and m. The axion energy, $\omega_a$, follows a Boltzmann distribution as defined in~Eq.~\eqref{eq:Boltzmann}. For simplicity, we have again used $\omega_a \sim m_a$. Defining $\cos\theta \equiv \hat{{\bf n}}_{\bf 1}  \cdot \hat{{\bf n}}_{\bf 2} $ and $\cos \theta' \equiv \hat{{\bf n}}_{\bf a} \cdot \hat{{\bf n}}_{\rm {\bf m}}$, energy and momentum conservation, solving for $\omega_1$, give
\begin{equation} \label{eq:omega1}
\begin{split}
\omega_1
&= \frac{\Omega_{\rm m}}{2}  \left(1 - \frac{m_a}{\Omega_{\rm m}}  \pm   \frac{1}{|\sin \frac{\theta}{2}|}
\sqrt{ \frac{1}{c_s^2}  - \cos^2 \frac{\theta}{2}   + \left( v^2   - \cos^2 \frac{\theta}{2} \right) \frac{m_a^2}{\Omega_{\rm m}^2}+ 2 \left(\cos^2 \frac{\theta}{2} - \frac{v}{c_s}  \cos\theta' \right) \frac{m_a}{\Omega_{\rm m}} }  \,
 \right).
\end{split}
\end{equation}
As the lasers are applied along the cavity axis perpendicular to the end mirrors, the photon states are driven to be one-directional.
In such a case, the $\theta$ angle can only take two values: $\theta =0$ or $\theta = \pi$. The photon momentum is unphysical, given a non-zero $\Omega_{\rm m}$, when $\theta = 0$. The only physical solution is $\theta = \pi$, {\it i.e.}~the incoming and outgoing photons are back-to-back. In this case, Eq.~\eqref{eq:omega1} simplifies to
\begin{equation} \label{eq:w11}
\begin{split}
\omega_1 
&= \frac{\Omega_{\rm m}}{2}  \left(1 - \frac{m_a}{\Omega_{\rm m}}  \pm  
\sqrt{ \frac{1}{c_s^2} +v^2 \frac{m_a^2}{\Omega_{\rm m}^2}-2  \frac{v}{c_s}  \cos\theta'  \frac{m_a}{\Omega_{\rm m}} }  \,
 \right).
\end{split}
\end{equation}
For the range of axion masses we are interested in, $m_a v \ll m_a \ll \Omega_{\rm m}\,c_s^{-1} \approx \omega_1 \sim $ eV, the term dependent on the angle $\theta'$ is always negligible. Therefore, we will assume $\theta' = 0$ in the following. Note that, with $c_s \ll 1$,~Eq.~\eqref{eq:w11} further simplifies to take the approximate form $\omega_1 \approx \Omega_{\rm m}/(2 c_s)$.
Fixing the angles as argued above, the energy and momentum conservation relations in~Eq.~\eqref{eq:cons-all} are satisfied exactly with
\begin{equation} \label{eq:cons-exact}
\begin{split}
\Omega_{\rm m} = c_s \frac{2  \omega_1 +  (1+v)m_a}{1+c_s},\quad  \text{ and } \quad \omega_2 = \frac{ (1-c_s ) \omega_1 + (1-v\,c_s)m_a}{1+c_s}.
\end{split}
\end{equation}
Note that we have used the continuous momentum conservation above. However, in a cavity, where the optical and acoustic modes are quantized, the momentum matching is parametrized by the mode overlap factor $a_{\rm ovl}$ given in~Eq.~\eqref{eq:Vdelta} (see also Eq.~\eqref{eq:aovl} in the main text).
To respect the kinematic matching condition, with fixed $\omega_1$ for a given $m_a$, $\omega_2$ and $\Omega_{\rm m}$ should be chosen to satisfy $a_{\rm ovl} \sim  1$. Hence, assuming the pump laser being on-resonance with the cavity,
\begin{equation}\label{eq:nm}
n_{\rm m} = n_1 + n_2 + \left[ \frac{v\, m_a}{\omega_{\rm FSR}} \right] = 2 n_1 + \left[\frac{m_a-\Omega_{\rm m}}{\omega_{\rm FSR}}\right] + \left [ \frac{v\, m_a}{\omega_{\rm FSR}} \right ] ,
\end{equation}
which is the cavity realization of Eq.~\eqref{eq:cons-exact} and perfectly agrees with it when the lasers and the mechanical drive are on-resonance. Above and hereafter, the square brackets denote the integer part of their argument. $\omega_{\rm FSR} \equiv \pi/L$ (free spectral range) defines the frequency separation between consecutive optical modes. Note that the last term in Eq.~\eqref{eq:nm}, coming from the axion momentum, is zero for $m_a < \text{meV}$ in a meter-long cavity. In the above equation and hereafter, we take $n_i$ to be the natural number labelling the mode $i$, {\it i.e.} $n_i = |i_z|$ where $i_z$ is defined in Eq.~\eqref{eq:quantizationp}. Eq.~\eqref{eq:sincmatch} in the main text is recovered taking into account that $\Omega_{\rm m} \sim \omega_{\rm FSR} \ll n_1 \, \omega_{\rm FSR}$.
Requiring $\omega_2$ to be a resonance mode frequency, Eq.~\eqref{eq:cons-exact} specifies how the cavity length must be tuned,
\begin{equation} \label{eq:modematch}
\begin{split}
n_1 - n_2 = \frac{L}{\pi} \frac{2 c_s \omega_{{\rm pump}} - (1- v\, c_s) m_a}{1+c_s}.
\end{split}
\end{equation}
where we have assumed the pump laser to be on-resonance, {\it i.e.} $\omega_1 = \omega_{\rm pump}$.
Note that, in the absence of an axion, the Stokes and anti-Stokes sidebands reside at $n_{\rm S,aS} \approx  n_1 \approx n_{\rm m}/2$. For example, for a meter-long cavity, $n_{\rm S,aS} =  n_1 \pm 2$ with $n_1 \sim \mathcal{O}(10^6) $. From~Eq.~\eqref{eq:modematch}, one observes that in the light axion regime ($m_a \lesssim  \omega_{\rm FSR}$), the axion sideband resides at the same mode as the Stokes sideband, while in the heavy regime ($m_a \gtrsim  \omega_{\rm FSR}$,) the axion sideband resides at a different mode. We note that, depending on the experimental strategy, $\omega_2$ can be slightly off-resonance, and the phase matching can be mildly violated with a suppressed $a_{\rm ovl}$.

\section{Derivation for rates of axion and optomechanical processes}
\label{sec:Srate}

We now derive rates for the relevant processes with the axioptomechanical and optomechanical couplings.
The optical and acoustic modes in the cavity can be driven by external photon lasers and a mechanical drive. The Hamiltonian terms describing the driven system read \cite{PhysRevA.30.1386,PhysRevA.31.3761,Aspelmeyer:2013lha,10.1093/oso/9780198828143.003.0005,RevModPhys.82.1155}
\begin{equation} \label{eq:}
\begin{split}
H_{\rm opt}
\supset \sum_n 
\omega_{n} a_n(t) a_n^{\dagger} (t) + \sum_n \sqrt{\kappa} \left( a_{\rm in} (t) a_n^{\dagger} (t) + {\rm h.c.} \right) + H_{\rm laser},
\end{split}
\end{equation}
for the optical modes, and similarly for the acoustic modes
\begin{equation} \label{eq:}
\begin{split}
H_{\rm ac} \supset \sum_k \Omega_{{\rm m},k} b_k(t) b_k^{\dagger} (t) + \sum_k  \sqrt{\Gamma_{\rm m}} \left( b_{\rm in} (t) b_k^{\dagger} (t) + {\rm h.c.} \right)+ H_{\rm mechanical~drive}.
\end{split}
\end{equation}
The sums run over integer numbers $n,k = 1,2,\cdots$, corresponding to the optical and acoustic cavity modes $a_n$ and $b_k$, respectively, with frequency
\begin{equation} \label{eq:}
\begin{split}
\omega_n = |{\bf p}_n|= n \frac{\pi}{L},\quad \text{ and } \quad \Omega_{{\rm m},k} = c_s|{\bf k}_{{\rm {\bf m}},k}| = k \frac{\pi c_s}{L}.
\end{split}
\end{equation}
Above, $\kappa$ and $\Gamma_{\rm m}$ are the optical and mechanical decay rates, respectively, given by the properties of the cavity and the filling material,
\begin{equation} \label{eq:}
\begin{split}
\kappa = \frac{ \pi}{\mathcal{F}_{\rm opt} L},\quad \text{ and } \quad
\Gamma_{\rm m} = \frac{c_s \pi}{\mathcal{F}_{\rm ac} L},
\end{split}
\end{equation}
where the optical and acoustic finesse of the cavity are determined by the reflectivity of the end mirrors, $r_{1,2}^{\rm (ac),(opt)}$, as follows
\begin{equation} \label{eq:}
\begin{split}
\mathcal{F}_{\rm ac,opt} = \frac{1}{\pi} \,  \frac{1-r^{\rm (ac),(opt)}_1r^{\rm (ac),(opt)}_2}{r^{\rm (ac),(opt)}_1 r^{\rm (ac),(opt)}_2}.
\end{split}
\end{equation}

The optical and acoustic modes in the cavity are coherent states, denoted as $| \alpha_i \rangle$ and $|\beta_i \rangle$, respectively, where the $i$ labels the cavity mode. These are eigenstates of the creation and annihilation operators,
\begin{equation} 
\begin{split}
& a_n | \alpha_n \rangle = \alpha_n | \alpha_n \rangle ,\qquad    \text{ where } \quad |\alpha_n|^2 = \langle \alpha_n | a_n^{\dagger} a_n  | \alpha_n \rangle = N_{\gamma,n}^{\rm circ}  , \\
& b_k | \beta_k \rangle = \beta_k | \beta_k \rangle ,\, \, \qquad \,\, \! \text{ where } \quad |\beta_k|^2 = \langle \beta_k | b_k^{\dagger} b_k  | \beta_{k} \rangle = N_{\phi,k}^{\rm circ} .
\end{split}
\end{equation}

The density of states of the driven optical and acoustic modes is determined by the photon lasers and the mechanical drive. For finite width laser input described by
\begin{equation} \label{eq:}
\begin{split}
a_{\rm in}(t) = \alpha_{\rm L} e^{-i\omega_{\rm L}t - \kappa_{\rm L} |t|/2},
\end{split}
\end{equation}
where $\omega_{\rm L}$ is the laser frequency and $\kappa_{\rm L}$ is the laser width, according to the input-output theorem, the optical density of states reads
\begin{equation} \label{eq:}
\begin{split}
\rho(\omega) =   \frac{1}{\pi} \frac{\kappa_{\rm L}/2}{(\omega - \omega_{\rm L})^2 + (\kappa_{\rm L}/2)^2},
\end{split}
\end{equation}
assuming $\kappa_{\rm L} \ll \kappa$. The total photon population number in the mode $n$ is related to the laser power $P_{\rm L}$ as
\begin{equation} \label{eq:ng}
\begin{split}
N_{\gamma,n}^{\rm circ} =  |\alpha_n|^2 = \frac{4 \, P_{\rm L}}{\omega_{\rm L} \kappa} \frac{(\kappa/2)^2}{(\omega_{\rm L} - \omega_{n})^2 + (\kappa/2)^2} \equiv N_{\gamma,{\rm L}} \frac{(\kappa/2)^2}{(\omega_{\rm L} - \omega_{n})^2 + (\kappa/2)^2}.
\end{split}
\end{equation}
The cavity mode with the frequency closest to the laser frequency has the largest population. Thus, we consider processes with only the closest resonant mode to the optical laser frequency. In the following, $N_{\gamma,{\rm pump}}^{\rm circ}$ ($N_{\gamma,{\rm probe}}^{\rm circ}$) will refer to the optical mode $n$ excited by the pump (probe) laser, {\it i.e.} the closest mode to the laser frequency. For the mechanical mode, similarly, the density of states reads
\begin{equation} \label{eq:}
\begin{split}
\rho_{\rm m}(\Omega) =   \frac{1}{\pi} \frac{\Gamma_{\rm L}/2}{(\Omega - \Omega_{\rm m})^2 + (\Gamma_{\rm L}/2)^2},
\end{split}
\end{equation}
with $\Omega_{\rm m}$ and $\Gamma_{\rm L}$ being the frequency and the width of the mechanical drive, respectively. The phonon population in the driven mode is also reduced from the total input if $\Omega_{\rm m} \ne \Omega_{{\rm m},k}$ according to 
\begin{equation} \label{eq:nphi}
\begin{split}
N_{\phi,k}^{\rm circ} =  |\beta_k|^2  = N_{\phi} \frac{(\Gamma_{\rm m}/2)^2}{(\Omega_{\rm m} - \Omega_{{\rm m},k})^2 + (\Gamma_{\rm m}/2)^2}.
\end{split}
\end{equation}
We do not specify how the input phonon population $N_{\phi}$ is related to the mechanical drive, instead, we take fixed values limited by various considerations to be discussed in the next section. As in the previous case, we only consider the acoustic resonant mode closest to the frequency of the coherent phonons and drop the index $k$ to simplify notation.

In the absence of an optical laser or mechanical drive, the optical and acoustic modes have density of states determined by the cavity decay rates
\begin{equation} \label{eq:}
\begin{split}
\rho_{n} (\omega) = 
\frac{1}{\pi}\frac{\kappa/2}{(\omega - \omega_{n})^2 + (\kappa/2)^2},\quad \text{ and } \quad 
\rho_{{\rm m}, k} (\Omega) = \frac{1}{\pi}
\frac{\Gamma_{\rm m}/2}{(\Omega - \Omega_{{\rm m}, k})^2 + (\Gamma_{\rm m}/2)^2}.
\end{split}
\end{equation}

We use Fermi's golden rule to derive the interaction rate. Note that, in general, Fermi Golden's rule is derived by assuming that the initial and final states are eigenstates of the bare Hamiltonian. Although coherent states are not, they are {\it amplitude}-eigenstates of the time evolution operator, {\it i.e.} $e^{i H_0 t} |\alpha \rangle = |\alpha e^{i \omega_i t} \rangle$, where $\omega_i$ is the frequency of the corresponding mode. The latter phase is responsible of the energy conservation delta function in Fermi's golden rule. 
Note that, as reasoned above, we only calculate the interaction rate for the initial optical mode with the closest resonance frequency $\omega_{n}$ to the pump laser frequency, {\it i.e.}~the mode number $n_\text{pump} = [ \omega_{\rm pump}/ \omega_{\rm FSR} ]$, and similarly for the final populated optical or acoustic modes, taken to be the closest to $\omega_{\rm probe}$ or $\Omega_{\rm m}$, respectively; $n_\text{probe} = [ \omega_{\rm probe}/ \omega_{\rm FSR} ]$, or $n_{\Omega_{\rm m}} = [\Omega_{\rm m}/(c_s \omega_{\rm FSR})]$. The corresponding detuning for each mode is defined as in the main text,
 \begin{equation}
 \Delta_{\rm pump,probe} \equiv \omega_{\rm pump,probe} - n_\text{pump,probe}\,  \omega_{\rm FSR}, \quad \text{ and } \quad \Delta_{\rm m} \equiv \Omega_{\rm m} - n_{\Omega_{\rm m}} \, \omega_{\rm FSR} \,c_s.
 \end{equation}
Treating the axion as a classical field, according to Fermi's golden rule, in the case of a populated final state of photons, the rate reads
\begin{equation}
\begin{split}
\Gamma 
&=\sum_{I,F} | \langle \alpha_{{\rm pump}}, \alpha_{{\rm probe}}, 1 | H_{\rm eff}  | \alpha_{{\rm pump}}, \alpha_{{\rm probe}}, 0 \rangle |^2 (2\pi) \delta ( \omega_{I} - \omega_F ) \\
&= (2\pi) \sum_{I,F} | \langle \alpha_{{\rm pump}}, \alpha_{{\rm probe}},1 | \sum_{i, j, k}  g_a^{(0)} \big({\bf p}_{i},{\bf p}_{j},{\bf k}_{{\rm {\bf m}},k} ,{\bf k_{\rm {\bf a}}} \big) g_{a\gamma\gamma} a(\omega_a) a_{i} a^{\dagger}_{j} b_k^\dagger |\alpha_{{\rm pump}}, \alpha_{{\rm probe}}, 0  \rangle |^2 \delta ( \omega_{I} - \omega_F ) \\
&=  (2\pi)  \sum_{I,F}  \left| g_a^{(0)} \right|^2  g_{a\gamma\gamma}^2 |a(\omega_a)|^2 
|\alpha_{{\rm pump}}|^2\, |\alpha_{{\rm probe}}|^2\, 
\delta (\omega_{I} - \omega_F) \\
&= (2\pi)  g_{a\gamma\gamma}^2  \left|  g_a^{(0)} \right|^2  \frac{2\rho_a}{m_a^2}  N_{\gamma,{\rm pump}}^{\rm circ}(\Delta_{\rm pump}) N_{\gamma,{\rm probe}}^{\rm circ}(\Delta_{\rm probe})
\\
&\qquad   \times \int d \omega_a \, d \omega_1 \, d \omega_2 \, d\Omega \, B_{m_a}(\omega_a) L(\omega_1-\omega_{\rm pump},\kappa_{\rm L}) L(\omega_2-\omega_{\rm probe},\kappa_{\rm L}) L(\Omega - \Omega_{n_{\rm m}}, \Gamma_{\rm m}) \delta (\omega_1 + \omega_a - \omega_2 - \Omega ) \\
& \simeq (2\pi)  g_{a\gamma\gamma}^2   \left|  g_a^{(0)} \right|^2  \frac{2\rho_a}{m_a^2} N_{\gamma,{\rm pump}}^{\rm circ}(\Delta_{\rm pump}) N_{\gamma,{\rm probe}}^{\rm circ}(\Delta_{\rm probe})  \\
& \qquad \qquad \qquad \quad  \  \ \qquad \times
\begin{cases}
 \int  d \omega_2 \, B_{m_a}(\omega_2 + \Omega_{n_{\rm m}} - \omega_{\rm pump})L(\omega_2-\omega_{\rm probe},\kappa_{\rm L}) ,\quad &\Delta f_a \gg \kappa_{\rm L} \\
 \int  d \omega_2  \, L(\omega_2 + \Omega_{n_{\rm m}} - m_a - \Delta f_a/2 -\omega_{\rm pump},\kappa_{\rm L}) L(\omega_2-\omega_{\rm probe},\kappa_{\rm L}) , \quad &\Delta f_a \ll \kappa_{\rm L}\\
  \end{cases} \\
& \simeq (2\pi)  g_{a\gamma\gamma}^2   \left| g_a^{(0)} \right|^2  \frac{2\rho_a}{m_a^2}  N_{\gamma, {\rm pump}}^{\rm circ}(\Delta_{\rm pump}) N_{\gamma, {\rm probe}}^{\rm circ}(\Delta_{\rm probe})
\begin{cases}
 \displaystyle B_{m_a}(\omega_{\rm probe} + \Omega_{n_{\rm m}} - \omega_{\rm pump}),\quad &\Delta f_a \gg \kappa_{\rm L}, \\
  \displaystyle L(\omega_{\rm probe} + \Omega_{n_{\rm m}} - \omega_{\rm pump}-m_a - \Delta f_a/2,2 \kappa_{\rm L})
  ,\quad &\Delta f_a \ll \kappa_{\rm L}, \\
  \end{cases}
\end{split}
\end{equation}
where we abbreviate a Lorentzian distribution with a full-width-at-half-maximum $\delta x$ and a central value $x_0$ as $L(x - x_0,\delta x)$, as displayed in Eq.~\eqref{eq:LorentzianStandard}. The Boltzmann distribution, $B_{m_a}(\omega_a)$, is given in Eq.~\eqref{eq:Boltzmann}. $\omega_1$ and $\omega_2$ are the frequencies of the initial and final photons, respectively, as defined in the main text.
Note that the sum over initial and final states, $\sum_{I,F}$, is non-trivial because of the finite width of the optical and acoustic modes. Above we have assumed, without loss of generality, that $\Gamma_{\rm m} \ll \kappa_L$. Similarly, in the case that the populated final state are the phonons by applying a mechanical drive with frequency $\Omega_{\rm m}$,

\begin{equation}
\begin{split}
\Gamma 
&=\sum_{I,F} | \langle \alpha_{{\rm pump}}, 1, \beta_{\Omega_{\rm m}} | H_{\rm eff}  | \alpha_{{\rm pump}}, 0, \beta_{\Omega_{\rm m}}  \rangle |^2 (2\pi) \delta ( \omega_{I} - \omega_F ) \\
&=  (2\pi)  \sum_{I,F}  \left| g_a^{(0)} \right|^2  g_{a\gamma\gamma}^2 |a(\omega_a)|^2 
|\alpha_{{\rm pump}}|^2\,  |\beta_{\Omega_{\rm m}}|^2
(2\pi) \delta (\omega_{I} - \omega_F) \\
&= (2\pi)  g_{a\gamma\gamma}^2  \left|  g_a^{(0)} \right|^2  \frac{2\rho_a}{m_a^2}  N_{\gamma,{\rm pump}}^{\rm circ}(\Delta_{\rm pump})  N_{\phi}^{\rm circ}(\Delta_{\rm m})
\\
&\qquad   \times \int d \omega_a \, d \omega_1 \, d \omega_2 \, d\Omega \, B_{m_a}(\omega_a) L(\omega_1-\omega_{\rm pump},\kappa_{\rm L}) L(\omega_2-\omega_{n_2},\kappa_{\rm L}) L(\Omega - \Omega_{\rm m}, \Gamma_{\rm L}) \delta (\omega_1 + \omega_a - \omega_2 - \Omega ) \\
& \simeq (2\pi)  g_{a\gamma\gamma}^2   \left|  g_a^{(0)} \right|^2  \frac{2\rho_a}{m_a^2} N_{\gamma,{\rm pump}}^{\rm circ}(\Delta_{\rm pump})N_{\phi}^{\rm circ}(\Delta_{\rm m}) \\
& \qquad \qquad \qquad \quad  \  \ \qquad \times
\begin{cases}
 \int  d \omega_2 \, B_{m_a}(\omega_2 + \Omega_{\rm m} - \omega_{\rm pump})L(\omega_2-\omega_{n_2},\kappa) ,\quad &\Delta f_a \gg \kappa_{\rm L} \\
 \int  d \omega_2  \, L(\omega_2 + \Omega_{\rm m} - m_a - \Delta f_a/2 -\omega_{\rm pump},\kappa_{\rm L}) L(\omega_2-\omega_{n_2},\kappa) , \quad &\Delta f_a \ll \kappa_{\rm L} \\
  \end{cases} \\
& \simeq (2\pi)  g_{a\gamma\gamma}^2   \left| g_a^{(0)} \right|^2  \frac{2\rho_a}{m_a^2}  N_{\gamma, {\rm pump}}^{\rm circ}(\Delta_{\rm pump})  N_{\phi}^{\rm circ}(\Delta_{\rm m})
\begin{cases}
 \displaystyle B_{m_a}(\omega_{n_2} + \Omega_{\rm m} - \omega_{\rm pump}),\quad &\Delta f_a \gg \kappa, \\
  \displaystyle L(\omega_{n_2} + \Omega_{\rm m} - \omega_{\rm pump}-m_a - \Delta f_a/2,\kappa)
  ,\quad &\Delta f_a \ll \kappa. \\
  \end{cases}
\end{split}
\end{equation}
Above, $\omega_{n_2}$ is the closest resonance frequency to the outgoing photon.
Note that we have assumed the laser and mechanical drive widths $\Gamma_{\rm L} \ll  \kappa_{\rm L} < \kappa$. 
Such spontaneously emitted photons from the axion sideband could be detected with a single photon detector. Note that, because the frequency of the phonon population is chosen to kinematically match the axion mass ($a_{\rm ovl} \sim 1$), the search is narrow-band in its nature up to $\omega_{\rm FSR}$, within which the momentum conservation can be violated in the cavity setup. 
Filters can be then employed to forbid most of the background photons from the transmitted
pump laser and the Stokes, anti-Stokes sidebands. Various types of filtering can be applied either before or after the cavity depending on explicit experimental strategies, which we leave for a future experimental proposal.

\section{Stokes and anti-Stokes processes and the phonon number limit}
\label{sec:SMphononN}

In this section, we derive the phonon number limit, and discuss the usual Stokes/anti-Stokes processes for reference. 
The importance of the Stokes/anti-Stokes sidebands depends on the detection modality, which we discuss for the two setups:

(a) {\it Final state populated with phonons.}  In such a case,  the Stokes/anti-Stokes sidebands are also enhanced by the phonon population. The Stokes rate reads
\begin{equation} 
\begin{split}
\Gamma_{\rm S} 
&=\sum_{I,F} | \langle \alpha_{\text{pump}}, 0, \beta_{\Omega_{\rm m}} | H_{\rm eff}  | \alpha_\text{pump}, 1, \beta_{\Omega_{\rm m}}  \rangle |^2 (2\pi) \delta ( \omega_{I} - \omega_F ) \\
&= (2\pi)  |g_0|^2  N_{\gamma,{\rm pump}}^{\rm circ} N_{\phi}^{\rm circ} \int d \omega_1 \, d \omega_2 \, d\Omega \,   L(\omega_1 - \omega_{\rm pump},\kappa_{\rm L}) L(\omega_2 - \omega_{n_{\rm S}},\kappa)  L(\Omega - \Omega_{\rm m},\Gamma_{\rm L}) \delta (\omega_1  - \omega_2 - \Omega ) \\
&\simeq (2\pi)  |g_0|^2  N_{\gamma,{\rm pump}}^{\rm circ} N_{\phi}^{\rm circ} \int  d \omega_2  \,   L(\omega_2 + \Omega_{\rm m}  - \omega_{\rm pump},\kappa_{\rm L}) L(\omega_2 - \omega_{n_{\rm S}},\kappa)   \\
&\simeq (2\pi)  |g_0|^2  N_{\gamma,{\rm pump}}^{\rm circ} N_{\phi}^{\rm circ}  \,  L(\omega_{\rm pump} -  \Omega_{\rm m}  - \omega_{n_{\rm S}},\kappa),   \\
\end{split}
\end{equation}
where $\omega_{n_{\rm S}}$ is the optical resonance mode that dominates the Stokes rate.
Similarly, for the anti-Stokes background,
\begin{equation} \label{eq:}
\begin{split}
\Gamma_{\rm aS} 
&= (2\pi)  |g_0|^2  N_{\gamma,{\rm pump}}^{\rm circ} N_{\phi}^{\rm circ} L(\omega_{\rm pump} + \Omega_{\rm m}  - \omega_{n_{\rm aS}},\kappa),  \\
\end{split}
\end{equation}
where $\omega_{n_{\rm aS}}$ is now the optical resonance mode that dominates the anti-Stokes rate. Tails of such Stokes/anti-Stokes sidebands within the axion signal window generically introduce a large photon count rate, similar to the tail of the pump laser. Such photon counts can however be suppressed by filtering the tails of the pump laser and the mechanical drive, as the~power spectral densities of the sidebands inherit the shape of the input drives. Additional filtering can be implemented noticing the polarization difference between such photons and the photons from the axion sideband. As the number of circulating phonons increases, one may enter the regime where $\Gamma_{\rm S,aS} \gtrsim P_{\rm pump}/\omega_{\rm opt}$, i.e.~the strong coupling regime~\cite{Aspelmeyer:2013lha} with an effective coupling $g \sim g_0 \sqrt{N^{\rm circ}_{\phi} /[1+(2\, \Delta_{\rm S,aS}/\kappa)^2]} \gtrsim \kappa$. $\Delta_{\rm S,aS}$ is the detuning between the (anti-)Stokes photon frequency and the closest optical mode. Notice that $g$ scales as $1/m_a$ when the phonon population is chosen to kinematically match the axion process. Thus, the heavier the axion mass, the later the strong coupling regime is reached. Higher-order optomechanical effects should be considered in such a regime, which may have relevant impact on the backgrounds, e.g.~higher-order sidebands, or the axion signal, e.g.~through pump depletion. We postpone such calculations to future work and allow for the strong coupling regime when choosing the number of circulating phonons.

$N_\phi^{\rm circ}$ is ultimately limited by the amount of acoustic energy $U_{\rm m}$ that the cavity can host,
\begin{equation} \label{eq:numberphononsmax}
\begin{split}
N_{\phi}^{\rm circ}= \frac{U_{\rm m}}{\Omega_{\rm m}},\quad \text{ and }\quad
U_{\rm m} \lesssim \frac{1}{2} \, \rho_{\rm He} \, c_s^2 \left(\frac{\delta \rho_{\rm He}}{\rho_{\rm He}}\right)^2 V_{\rm mode}.
\end{split}
\end{equation}
Imposing $\delta \rho_{\rm He} / \rho_{\rm He} < 10^{-3}$, the number of phonons should not surpass $N_{\phi}^{\rm circ} \lesssim 10^{19} (L/{\rm m})^2$, assuming $\lambda_{\rm pump} = 1.5 \, \mu\text{m}$.

(b) {\it Final state populated with photons.}
In principle, as the photons from the probe laser should be selected to have a different polarization than the photons from the pump laser, the probe laser does not populate the final states participating in the Stokes and anti-Stokes processes. However, experimentally it may be challenging to control the polarization of all photons in the probe laser.  
Assuming a percentage $\epsilon_{\rm pol}$ of the photons in the probe laser that enjoy the proper polarization to quantum-mechanically enhance the Stokes process, the rate is given by
\begin{equation} 
\begin{split}
\Gamma_{\rm S} 
&=\sum_{I,F} | \langle \alpha_{\text{pump}}, \alpha_{\rm probe}^{\rm pol}, 1 | H_{\rm eff}  | \alpha_\text{pump}, \alpha_{\rm probe}^{\rm pol}, 0 \rangle |^2 (2\pi) \delta ( \omega_{I} - \omega_F ) \\
&= (2\pi)  |g_0|^2  N_{\gamma,{\rm pump}}^{\rm circ} ( \epsilon_{\rm pol}  N_{\gamma,{\rm probe}}^{\rm circ} ) \int d \omega_1 \, d \omega_2 \, d\Omega \,   L(\omega_1 - \omega_{\rm pump},\kappa_{\rm L}) L(\omega_2 - \omega_{\rm probe},\kappa_{\rm L})  L(\Omega - \Omega_{n'_{\rm m}},\Gamma_{\rm m}) \delta (\omega_1  - \omega_2 - \Omega ) \\
&\simeq (2\pi)  |g_0|^2  N_{\gamma,{\rm pump}}^{\rm circ} ( \epsilon_{\rm pol}  N_{\gamma,{\rm probe}}^{\rm circ} ) \int d \omega_2 \,  L( \omega_2 + \Omega_{n'_{\rm m}}  - \omega_{\rm pump},\kappa_{\rm L}) L(\omega_2 - \omega_{\rm probe},\kappa_{\rm L})   \\
&\simeq \epsilon_{\rm pol} \, (2\pi)  |g_0|^2  N_{\gamma,{\rm pump}}^{\rm circ} N_{\gamma,{\rm probe}}^{\rm circ}   \,  L(\omega_{\rm probe}+\Omega_{n_{\rm m}'}-\omega_{\rm pump},2 \kappa_{\rm L}),   \\
\end{split}
\label{eq:pumpdepprobe}
\end{equation}
and similarly for the anti-Stokes process $\Gamma_{\rm aS} \simeq \epsilon_{\rm pol} \, (2\pi)  |g_0|^2  N_{\gamma,{\rm pump}}^{\rm circ} N_{\gamma,{\rm probe}}^{\rm circ}   \,  L(m_a -\Omega_{\rm m} - \Omega_{n'_{\rm m}},2 \kappa_{\rm L})$. Considering the process with $a_{\rm ovl} \sim 1$, the rate is suppressed by $L(m_a,2\kappa_L)$. Depending on the polarization control of the laser beam, within the axion signal window, background photons from such processes are sub-dominant compared to the probe laser for $m_a \gtrsim \sqrt{\epsilon_{\rm pol} \, N_{\gamma,{\rm pump}}^{\rm circ}}\, |g_{0}| $. One may also enter the strong coupling regime when the effective coupling $g \sim g_0 \epsilon_{\rm pol}\sqrt{N^{\rm circ}_{\gamma,{\rm L}} /[1+(\Delta_{\rm S,aS}/\kappa_{\rm L})^2]} \gtrsim \kappa$. We leave the study of higher-order effects to a future work and do not restrict the laser powers in the chosen benchmarks.

\section{Scalings of the sensitivity curves with the input parameters}
\label{sec:e}

In this section we display the explicit scaling of the $g_{a\gamma \gamma}$ vs $m_a$ sensitivity curves as a function of the experimental input factors. These factors are of four kinds: parameters related with the geometry of the cavity (length $L$ and optical finesse ${\cal F}_\text{opt}$), the material used to fill it (density $\rho$, speed of sound $c_s$, relative permittivity $\epsilon_r$), the photon lasers and the mechanical drive ($N_\phi$, $\kappa_L$, $P_\text{pump}$ and $P_\text{probe}$), and the axion parameters ($\rho_a$ and $m_a$). 

First, we discuss specifics of the scanning strategy to cover an extended axion mass range, that defines the reach to $g_{a\gamma\gamma}$ over a given integration time per each scan, $t_{\rm int}$, and the coverage in axion masses for a total exposure time of the experiment, $T_{\rm exp}$. As has been discussed in the main text, the reach curve is determined by two subsequent measurements,
separated by a probe / filter frequency window $\delta$.
The generic scanning strategy consists of matching $\delta$ to the relevant width of $\Gamma_\text{sig}$, up to an order one parameter $\epsilon$, in order not to lose on sensitivity:
\begin{enumerate}[(a)]
\item  For $\Delta f_a <2\, \kappa_L \text{ or} <\kappa$, {\it i.e.} $m_a \lesssim 5 \text{ neV} (\kappa_L/\text{Hz})$ or $\lesssim 74 \text{ neV} (\tfrac{10^7}{{\cal F}_{\rm opt}/\pi} \tfrac{1 \text{ m}}{L})$, the scanning step is set by $\delta = \epsilon \kappa_L$ or $=\epsilon \kappa / 2$, depending whether the populated final state are photons or phonons, respectively.
Two successive measurements overlap at $\delta/2 \approx \Delta + \Omega_{\rm m} - m_a $,  where 
\begin{equation}
{\rm SNR}_{\gamma \text{-pop}} \propto L(\delta/2 , 2 \kappa_{\rm L}) = \frac{1}{\pi \kappa_{\rm L}}\frac{1}{\epsilon^2/4+1}, \quad \text{or} \quad \text{SNR}_{\phi \text{-pop}} \propto L(\delta/2,\kappa) = \frac{2}{\pi \kappa}\frac{1}{\epsilon^2/4 + 1}.
\end{equation}
The coverage in axion masses in this regime grows linearly with the exposure time of the experiment as 
\begin{equation}
\text{coverage} = \delta\, \frac{T_{\rm exp}}{t_{\rm int}},
\end{equation}
 which for $\epsilon =1$ allows one to scan $\sim 20 \text{ neV/year}$ ($\sim 20 \text{ peV/year}$) for $\kappa_{L},\kappa/2 = {\rm 1 \, Hz}$ and $t_\text{int} = 1\, \text{s}$ $(=10^3 \, \text{s})$.
\item If $\Delta f_a > 2\,\kappa_L$ ($> \kappa$) in the photon (phonon) populated case, the width of the axion Boltzmann distribution sets the scanning step, $\delta
= \epsilon \, m_a v^2  / (4\pi)$. Separating consecutive scans by a frequency $\delta$, two successive measurements overlap at the condition $ \delta/(e^{2 \delta / \Delta f_a } - 1) \approx \Delta + \Omega_{\rm m} - m_a$.  
Hence, at the intersection it follows 
\begin{equation}
{\rm SNR}\propto B_{m_a}( \Delta + \Omega_{\rm m}  ) = \frac{1}{\Delta f_a} \left(\frac{2}{\pi} \left(\frac{\epsilon}{e^{\epsilon}-1} \right) e^{\displaystyle \epsilon/(e^{\epsilon}-1)}\right)^{\frac{1}{2}}.
\end{equation}
The coverage in axion masses over an experimental time $T_{\rm exp}$ is given by
\begin{equation}
\text{coverage} = m_a^{(0)} \left(
\big(\epsilon
\frac{ v^2}{4\pi}
+1 \big)^{\frac{T_\text{exp}}{t_\text{int}}}-1\right),
\end{equation}
where $m_a^{(0)}$ is the axion mass that peaks the sensitivity of the initial measurement of the scan. Taking $\epsilon = 1$, one can cover $ \sim 70 \, m_a^{(0)}$, which is almost two orders of magnitude in axion mass per year for $t_\text{int} = 1 \, \text{s}$, or $40\%$ of the axion mass per year if $t_\text{int} = 10^2 \, \text{s}$ and 10 cavities are simultaneously employed.\end{enumerate}

Having specified the choice of the scanning spacing $\delta$, and assuming that $t_\text{int} \geq \kappa_L^{-1} > \kappa^{-1}$, for the scenario with the populated final state photons ($\gamma-{\rm pop}$), 
\begin{equation}
g_{a\gamma \gamma}^{\gamma-{\rm pop}} \! \propto \!  \left(\rho^{1/2} c_s^{1/2} \frac{\epsilon_r +2}{\epsilon_r-1}\epsilon_r^{3/4}\right) \!\! \left(\frac{1}{{\cal F}_\text{opt}^{5/4}} \frac{1}{L^{1/4}}\right) \!\!\left( \frac{1}{\omega_\text{opt}^{5/4}} \frac{1}{P_\text{pump}^{1/2}}\frac{1}{P_\text{probe}^{1/4}} \right) \!\! \frac{m_a }{\rho_a^{1/2}} \times
\begin{cases}
m_a^{\tfrac{1}{2}} \left(\frac{e^\epsilon-1}{\displaystyle e^{\frac{\epsilon}{1-e^\epsilon}}\epsilon}\right)^{\tfrac{1}{4}} \left(\frac{1}{t_\text{int} \kappa_L}\right)^{\tfrac{1}{4}}\!\!\!\!\!, & t_\text{int} > (2\kappa_L)^{-1}\! > \tau_a,\\
\kappa_L^{\tfrac{1}{2}}(1+\epsilon^2/4)^{\tfrac{1}{2}} \left(\frac{1}{t_\text{int} m_a}\right)^{\tfrac{1}{4}} \!\!\!\!\!,& t_\text{int} > \tau_a >  (2\kappa_L)^{-1}\!\!\!\!\! ,\\
\kappa_L^{\tfrac{1}{2}}(1+\epsilon^2/4)^{\tfrac{1}{2}} \left(\frac{1}{t_\text{int} m_a}\right)^{\tfrac{1}{2}} \!\!\!\!\!, &  t_\text{int} < \tau_a,\\
\end{cases}
\end{equation}
while for the populated phonons in the final state ($\phi-{\rm pop}$), 
\begin{equation}
g_{a\gamma \gamma}^{\phi-{\rm pop}} \! \propto \!  \left(\rho^{1/2} c_s^{1/2} \frac{\epsilon_r +2}{\epsilon_r-1}\epsilon_r^{3/4}\right) \!\! \left(\frac{L^{1/2}}{{\cal F}_\text{opt}^{1/2}}\right)\!\!\left( \frac{1}{\omega_\text{opt}^{3/2}} \frac{1}{P_\text{pump}^{1/2}} \frac{1}{N_\phi^{1/2}} \right) \!\! \frac{m_a }{\rho_a^{1/2}} \times
\begin{cases}
m_a^{\tfrac{1}{2}} \left(\frac{e^\epsilon-1}{\displaystyle e^{\frac{\epsilon}{1-e^\epsilon}}\epsilon}\right)^{\tfrac{1}{4}} \!, & t_\text{int} > \kappa^{-1}\! > \tau_a,\\
({\cal F}_\text{opt}L)^{-\tfrac{1}{2}}(1+\epsilon^2/4)^{\tfrac{1}{2}}\! ,& t_\text{int} > \tau_a >  \kappa^{-1}\!\!\! ,\\
({\cal F}_\text{opt}L)^{-\tfrac{1}{2}}(1+\epsilon^2/4)^{\tfrac{1}{2}} \left(\frac{1}{t_\text{int} m_a}\right)^{\tfrac{1}{2}} \!\!\!\!\!, &  t_\text{int} < \tau_a,\\
\end{cases}
\end{equation}
Therefore, the sensitivity in the reach plots will scale with the axion mass, depending on how the integration time compares to the axion and laser coherence times, as follows
\begin{equation}
    g_{a\gamma \gamma}^{\gamma-{\rm pop}} \propto \begin{cases} m_a^{3/2}, &t_\text{int} >  (2\kappa_L)^{-1}>\tau_a ,\\
    m_a^{3/4}, & t_\text{int} > \tau_a >  (2\kappa_L)^{-1} ,\\
    m_a^{1/2}, &t_\text{int} < \tau_a  ,\\
    \end{cases} \qquad \text{ and } \qquad g_{a\gamma \gamma}^{\phi-{\rm pop}} \propto \begin{cases} m_a^{3/2}, &t_\text{int} > \kappa^{-1} > \tau_a, \\ m_a , & t_\text{int} > \tau_a > \kappa^{-1}, \\ m_a^{1/2}, & t_\text{int} < \tau_a. \end{cases}
\end{equation}
Note that, for the phonon populated case, in the heavy axion regime, saturating the number of phonons that the cavity can host which is limited by the density perturbations in the material as stated in Eq.~\eqref{eq:numberphononsmax},
\begin{equation}
N_\phi^{\rm max} \propto  \left(\frac{\rho}{\sqrt{\epsilon_r}} c_s \left( \frac{\delta \rho}{\rho}\right)^2\right)  \left(\frac{L^2}{\omega_{\rm opt}^2}\right),
\end{equation}
the sensitivity scales with $L$ as $g_{a\gamma \gamma} \propto L^{-1/2}$, so that larger cavities still lead to better sensitivities, although the impact of the length is milder than in the photon populated case. Remarkably, the sensitivity in the saturated phonon case does not depend on the speed of sound and density of the filling material. 
\end{document}